\documentclass[lettersize,journal]{IEEEtran}

\usepackage{amsmath,amssymb,amsfonts}
\usepackage{algorithm}
\usepackage{array}
\usepackage{textcomp}
\usepackage{stfloats}
\usepackage{url}
\usepackage{verbatim}
\usepackage{graphicx}
\usepackage{cite}
\usepackage{xcolor}
\usepackage{adjustbox}
\usepackage{algorithm}
\usepackage{algpseudocode}
\usepackage{cleveref}
\usepackage{multicol}
\usepackage{multirow}
\usepackage{array}
\usepackage[inline]{enumitem}
\usepackage{multirow}
\usepackage{float}
\usepackage{amsthm} 
\usepackage{hhline}
\usepackage{makecell}
\usepackage{MnSymbol,wasysym, marvosym} 

\usepackage[caption=false]{subfig}

\makeatletter
\def\BState{\State\hskip-\ALG@thistlm}
\makeatother

\newcommand{\algorithmicinput}{\textbf{Input:}}
\newcommand{\INPUT}{\item[\algorithmicinput]}
\newcommand{\algorithmicoutput}{\textbf{Output:}}
\newcommand{\OUTPUT}{\item[\algorithmicoutput]}

\newcommand{\ve}[1]{\boldsymbol{#1}}         
\newcommand{\ma}[1]{\boldsymbol{\rm{#1}}}    
\newcommand{\shape}[3]{\in\mathbb{#1}^{#2\times #3}}
\newcommand{\Hra}{{\boldsymbol{\rm{H}}_{ra}^{f_L}}}
\newcommand{\Hur}{{\boldsymbol{\rm{H}}_{ur}^{f_H}}}
\newcommand{\Hua}{{\boldsymbol{\rm{H}}_{ua}^{f_L}}}
\newcommand{\Wa}{{\boldsymbol{\rm{W}}_{a}^{f_L}}}
\newcommand{\Wur}{{\boldsymbol{\rm{W}}_{ur}^{f_H}}}
\newcommand{\Wua}{{\boldsymbol{\rm{W}}_{ua}^{f_L}}}
\newcommand{\Pur}{{P_{ur}^{f_H}}}
\newcommand{\Pua}{{P_{ua}^{f_L}}}

\newcommand{\ra}[1]{{\boldsymbol{\rm{#1}}_{ra}^{f_L}}}
\newcommand{\ur}[1]{{\boldsymbol{\rm{#1}}_{ur}^{f_H}}}
\newcommand{\ua}[1]{{\boldsymbol{\rm{#1}}_{ua}^{f_L}}}
\newcommand{\underPsi}[1]{{\boldsymbol{\rm{#1}}_{\ma{\Psi}}}}
\newcommand{\AlgoName}[1]{{\textbf{\emph{#1}}}}

\newtheorem{theorem}{Theorem}

\begin{document}

\title{Relay-Assisted Carrier Aggregation (RACA) Uplink System for Enhancing Data Rate of Extended Reality (XR)}

\author{Chi-Wei Chen,\IEEEmembership{~Student Member,~IEEE}, Wen-Chiao Tsai,\IEEEmembership{~Student Member,~IEEE}, Lung-Sheng Tsai, and An-Yeu (Andy) Wu,\IEEEmembership{~Fellow,~IEEE}.
        
\thanks{This work was ﬁnancially supported in part by the Ministry of Science and Technology of Taiwan under Grant MOST 110-2221-E-002-184-MY3, and in part by MediaTek, Inc., Hsinchu, Taiwan, under Grant MTKC-2022-0125 (Corresponding author: An-Yeu Wu)}

\thanks{Chi-Wei Chen, Wen-Chiao Tsai, and An-Yeu (Andy) Wu are with the Graduate Institute of Electronics Engineering and Department of Electrical Engineering, National Taiwan University, Taipei, 10617, Taiwan (e-mail: wilbur@access.ee.ntu.edu.tw; daniel@access.ee.ntu.edu.tw; andywu@ntu.edu.tw).}

\thanks{Lung-Sheng Tsai is with MediaTek, Inc., Hsinchu, 30078, Taiwan (e-mail: Longson.Tsai@mediatek.com).}

\vspace{-0.5em}
}

\maketitle
\begin{abstract}
In Extended Reality (XR) applications, high data rates and low latency are crucial for immersive experiences. Uplink transmission in XR is challenging due to the limited antennas and power of lightweight XR devices. To improve data transmission rates, we investigate a relay-assisted carrier aggregation (RACA) system. The XR device simultaneously transmits data to an access point (AP) and a relay in proximity over low-frequency and high-frequency bands, respectively. Then, the relay down-converts and amplifies the signals to the AP, effectively acting as an additional transmit antenna for the XR device. In this paper, we propose two algorithms to maximize the data rate of the XR device in their respective protocols. In the centralized protocol, the rate maximization problem is equivalently transformed as a weighted mean square error minimization (WMMSE) problem which can be solved iteratively by alternative optimization. In the distributed protocol, the rate maximization problem is decomposed into two independent sub-problems where the rate of the direct link and the rate of the relay link are maximized by singular value decomposition (SVD)-based methods with water-filling (WF). Simulation results show that the rate of the RACA system is improved by $32\%$ compared to that of the conventional carrier aggregation scheme.

\end{abstract}


\section{Introduction}
\label{sec:intro}
\IEEEPARstart{R}{ecently}, Extended Reality (XR) and cloud gaming applications have gained significant attention in 5G usage scenarios due to their high data rates and low latency requirements in wireless communications. XR encompasses Virtual Reality (VR), Augmented Reality (AR), and Mixed Reality (MR), offering immersive experiences across various domains such as entertainment, education, healthcare, and manufacturing \cite{ref:XR_case}. To realize XR applications, the achievement of high data rates is essential \cite{ref:XR_split,ref:XR_case}. Two straightforward methods to boost data rates are: 1) employing the massive multi-input multi-output (MIMO) technique to support the parallel transmission of multiple data streams, and 2) increasing transmit power to enhance the signal-to-noise ratio (SNR). However, portable user equipment (UE)-side devices, such as XR glasses, usually have limited number of antennas. It becomes the bottleneck for MIMO gain improvement. Moreover, UE-side devices also have limited transmit power due to battery constraints \cite{ref:XR_UE_power}. Therefore, in uplink transmission, the resource-limited UE requires other techniques to enhance data rates.

\begin{figure*}[tbp]
\centerline{\includegraphics[width=1\linewidth]{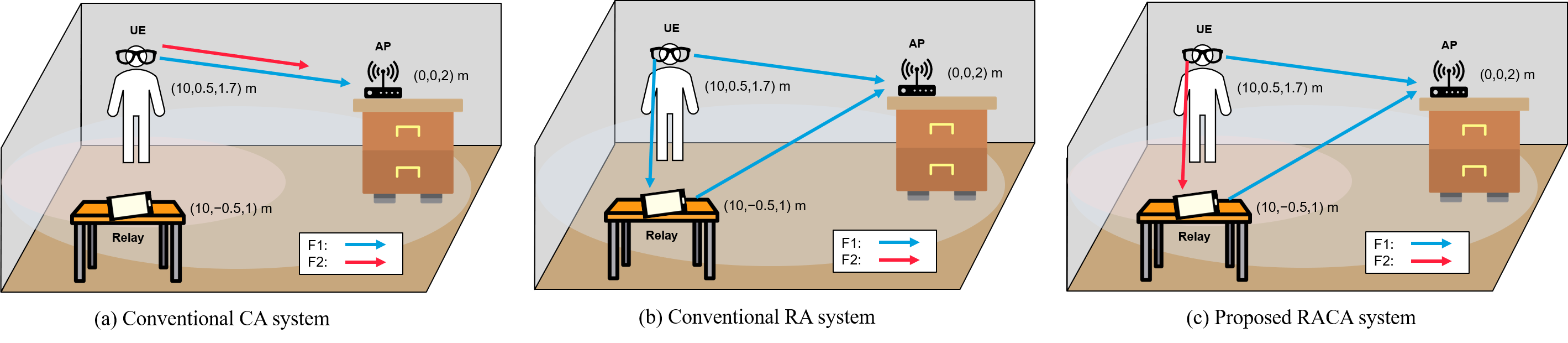}}
\vspace{-1em}
\caption{MIMO communication systems. (a) Conventional CA system, (b) Conventional RA system, (c) Proposed RACA system.}
\label{fig:system}
\end{figure*}

As shown in Fig. \ref{fig:system}(a), the carrier aggregation (CA) technique proposed by the Third Generation Partnership Project (3GPP) in Long Term Evolution (LTE)-Advanced standard can achieve data rate improvement via frequency-division multiplexing \cite{ref:CA_3gpp}. By aggregating multiple component carriers (CC) in multiple frequency bands, CA allows for wider overall bandwidth and higher data transmission speeds. However, due to the limited frequency spectrum resource, some CCs may be located in higher frequency bands and suffer from severe path loss. The challenges of high path loss and sensitivity to blockage in high frequency limit the coverage in practical applications.

On the other hand, to increase the SNR at a battery-operated UE with limited transmit power, the amplify-and-forward (AF) relay was introduced as a cooperative node to extend the coverage and improve the signal quality \cite{ref:relay_tutorial,ref:relay_mse,ref:relay_wmmse,ref:relay_hybrid}. Fortunately, most people may carry more than one device nowadays. Therefore, the secondary device, such as a smartphone, can act as an AF relay to support the primary UE, such as XR glasses, thus achieving device collaboration. The relay-assisted (RA) system is shown in Fig. \ref{fig:system}(b). Although AF relay can enhance SNR for better coverage, the limited antenna capability at the UE still bounds the scaling gain of MIMO capacity.

In summary, both the conventional CA and RA systems have their inherent limitations. The CA system is constrained by limited coverage at higher frequency bands, whereas the RA system suffers from a restricted MIMO rank. These limitations pose significant challenges in fulfilling the stringent data rate requirement of XR applications on lightweight UE-side devices. To address these issues, this paper investigates a relay-assisted carrier aggregation (RACA) cooperative system. The RACA system combines the benefits of multiplexing gain and coverage enhancement in CA and RA systems, respectively. More specifically, as illustrated in Fig. \ref{fig:system}(c), the primary UE concurrently transmits data to the access point (AP) and the nearby AF relay through $f_L$ and $f_H$ frequency bands, respectively. Transmitting data in different frequency bands exploits the frequency-division multiplexing as the CA system. Subsequently, after the frequency-translation AF relay receives the signal in the $f_H$ frequency band, the signal is down-converted to the $f_L$ frequency band, followed by the amplification and forwarding to the AP. Consequently, the AF relay improves the SNR and coverage as in the RA system. That is, multiple data streams are transmitted using frequency-division multiplexing at the antenna-limited primary UE, and received in spatial multiplexing at the antenna-abundant AP. In essence, the collaborative AF relay serves as a virtual antenna array for the primary UE while remaining transparent to the AP. The comparisons of different MIMO systems are summarized in Table \ref{table_system}. 

While the frequency-translation AF relay has been previously proposed to assist the antenna-limited UE in \cite{ref:SUDAS2015,ref:SUDAS2016,ref:UE-CoMIMO}, there are two issues should be investigated and addressed:

\begin{enumerate}[leftmargin=*]
    \item \emph{Spectral inefficiency in single-antenna relay systems without direct link}: The shared UE-side distributed antenna system (SUDAS) in \cite{ref:SUDAS2015}, \cite{ref:SUDAS2016} focuses on the use of multiple single-antenna relays, which only exploits frequency division multiplexing to enhance data rate rather than spatial multiplexing. On the other hand, the problem formulation neglecting the direct link makes the optimization more tractable but limits the spectral efficiency. 
    \item \emph{Lack of mathematical derivation as a theoretical benchmark for RACA system}: While the end-user-centric collaborative MIMO (UE-CoMIMO) system in \cite{ref:UE-CoMIMO} presents improvements with AF-relays in diversity, rank, and localization through system-level simulation, the comprehensive mathematical problem formulation and the dedicated optimization algorithms are not derived in the paper. Furthermore, the results of different technical aspects are not included to evaluate the system's effectiveness. 
\end{enumerate}

By considering a multiple-antenna relay-assisted system with a direct link, we propose two optimization algorithms in two proposed transmission protocols to maximize the sum rate and enhance spectral efficiency through spatial multiplexing. Our main contributions are summarized as follows:

\begin{table*}[tbp]
    \centering
    \renewcommand\arraystretch{1.5}
    \vspace{-0.5em}
    \caption{Comparison of Different MIMO Systems}
    \vspace{-1em}
    \begin{tabular}{|c|c|c|c|}
        \hline
          & \textbf{Conventional CA System} \cite{ref:CA_2010,ref:CA_3gpp,ref:CA_2012} & \textbf{Conventional RA System} \cite{ref:relay_tutorial,ref:relay_mse,ref:relay_wmmse,ref:relay_hybrid} & \textbf{Proposed RACA System}\\
        \hhline{|====|}
        \textbf{SNR Enhancement} & \raisebox{-0.5mm}{\includegraphics[width=3mm, height=3mm]{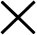}} & \raisebox{-0.5mm}{\includegraphics[width=3mm, height=3mm]{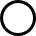}} & \raisebox{-0.5mm}{\includegraphics[width=3mm, height=3mm]{Figure/smile.png}} \\
        \hline
        \textbf{Multiplexing Gain} & \raisebox{-0.5mm}{\includegraphics[width=3mm, height=3mm]{Figure/smile.png}} & \raisebox{-0.5mm}{\includegraphics[width=3mm, height=3mm]{Figure/unsmile.png}} & \raisebox{-0.5mm}{\includegraphics[width=3mm, height=3mm]{Figure/smile.png}} \\
        \hline
        \textbf{Summary} & \quad Boosted data rate but short coverage \quad & \quad Enhanced SNR but still limited rank \quad & Combine the advantages of CA and RA \\
        \hline
    \end{tabular}
    \label{table_system}
    \vspace{-1em}
\end{table*}

\begin{enumerate}[leftmargin=*]
    \item \emph{Mathematical formulation and optimization of Relay-assisted carrier aggregation (RACA) system}: This is the first work to mathematically formulate the rate maximization problem and derive the optimization algorithms for the RACA system, aiming to simultaneously leverage the SNR enhancement of RA and the frequency-division multiplexing of CA. Furthermore, by considering the overhead of the estimated channel exchanges and the optimized beamformer exchanges, we propose centralized and distributed transmission protocols for various scenarios.   
    \item \emph{WMMSE-based optimization in the centralized system}: In the centralized protocol system, all estimated channels are centrally collected at the AP to perform the optimization. To tackle the non-convex rate maximization in the RACA system caused by the interference between the direct link and the relay link, we equivalently transform the intractable rate maximization problem into a weighted mean square error minimization (WMMSE) problem and solve it through alternative optimization. This convergence-guaranteed algorithm serves as an upper benchmark in terms of the sum rate for the RACA system.
    \item \emph{SVD-based optimization in the distributed system}: In the distributed protocol system, the optimization process is distributively offloaded to the UE and relay nodes with limited exchanges of channel state information (CSI). The rate maximization problem is decomposed into two independent sub-problems: maximization of the rate of direct link and the rate of relay link. We propose an efficient singular value decomposition (SVD) with the water-filling (WF) method as a practical solution, achieving $96\%$ of the centralized WMMSE algorithm’s performance with significantly lower complexity and reduced signal exchange overhead.
    \item \emph{Evaluation of the RACA system in different aspects}: To demonstrate the efficiency of the RACA system, we conduct extensive simulations to evaluate its performance across varying SNRs, frequency band combinations, and power allocation strategies. Simulation results show that the rate of our RACA system is improved by $32\%$ and $50\%$ compared to that of the CA and RA systems, respectively. 
\end{enumerate}

The rest of this paper is organized as follows. Section II reviews the prior works in different MIMO systems. Then, Section III introduces the RACA system model and problem formulation. Section IV presents the two transmission protocols. In Section V and Section VI, the WMMSE and SVD-WF optimization are derived in detail, respectively. The numerical results and analysis are shown in Section VII. Finally, Section VIII concludes our work.

\textit{Notations}: $a$, $\ve{a}$ and $\ma{A}$ denote scalar, vector and matrix, respectively. $\ma{A}^T, \ma{A}^H, \ma{A}^{-1}, \text{tr}(\ma{A}), |\ma{A}|$ are the transpose, conjugate transpose, inverse, trace, and determinant of matrix $\ma{A}$, respectively. $[\ma{A}]_{i,j}$ denotes the $(i,j)$-entry of matrix $\ma{A}$, and $\text{diag}(\ma{A})$ returns its diagonal vector. $\ma{I}_N$ is an $N\times N$ identity matrix. $\mathcal{CN}(\ve{a},\ma{A})$ denotes a circularly symmetric complex Gaussian distribution with mean $\ve{a}$ and covariance matrix $\ma{A}$. The expectation operator is denoted by $\mathbb{E}[\cdot]$. $\lVert \cdot \rVert$ is a norm operator. $\text{Re}\{a\}$ is a real part of $a$, and $[a]_+=\max\{0,a\}$. Note that $\mathbb{C}$ and $\mathbb{R}_+$ represent the sets of complex and nonnegative real numbers, respectively.

\section{Preliminaries and Priors Works}
\subsection{Conventional Carrier Aggregation (CA) Systems}
Utilizing wider transmission bandwidth to improve data rates via frequency-division multiplexing is an efficient technique for UE-side devices because data rates cannot be continuously increased by the multi-antenna technique due to the limited size and complexity constraints on UE-side devices \cite{ref:CA_2010}. In practice, it is hard to access a broad and continuous available spectrum. To utilize bandwidths more flexibly, 3GPP proposed a CA technique to combine multiple CC in LTE-Advanced standard \cite{ref:CA_3gpp}. By aggregating multiple frequency bands, CA allows for a wider overall bandwidth, resulting in higher data transmission rate and better network efficiency, as illustrated in Fig. \ref{fig:system}(a). Specifically, CA can be categorized into intra-band contiguous, intra-band non-contiguous and inter-band non-contiguous carrier aggregation \cite{ref:CA_2012}. To fully utilize a non-continuous spectrum, inter-band non-contiguous CA provides the highest flexibility to exploit different frequency bands for heterogeneous networks. It can even aggregate with abundant bandwidth resources in the millimeter wave (mmWave) range. While mmWave offers a broad bandwidth in the high-frequency spectrum, its coverage in practical applications is restricted by the severe propagation attenuation and its vulnerability to blockage \cite{ref:mmWave}.

\subsection{Conventional Relay-assisted (RA) Systems}
To increase the SNR with limited transmit power at battery-operated UE and combat severe propagation loss, especially in mmWave, the AF relay was introduced as a cooperative node to extend coverage and improve signal quality \cite{ref:relay_tutorial,ref:relay_mse,ref:relay_wmmse,ref:relay_hybrid}. The AF relay amplifies the received signal from the source, and forwards the amplified signal to the destination. In \cite{ref:relay_wmmse}, the precoders and relay amplifying matrix are jointly optimized to maximize the sum rate of multiple users in the mmWave scenarios. Furthermore, the hybrid beamforming in mmWave is exploited by a large-antenna AF relay to reduce implementation cost and complexity in \cite{ref:relay_hybrid}. However, the AF relay usually improves significantly when the direct link is weak or blocked. For generality, we consider the RA systems with the direct link shown as Fig. \ref{fig:system}(b). While the AF relay is capable of enhancing SNR to address the short-coverage challenges, the UE's limited antenna capability remains a bottleneck for the MIMO scaling gain.

\subsection{Relay-assisted Carrier Aggregation (RACA) Systems}
Different from the conventional in-band AF relay to receive and forward the signal at the same frequency band to enhance coverage, \cite{ref:SUDAS2015,ref:SUDAS2016,ref:UE-CoMIMO} aim to create the virtual antenna at the UE-side to augment the MIMO gain as the RACA system. In particular, \cite{ref:SUDAS2015,ref:SUDAS2016} introduced the concept of a shared UE-side distributed antenna system (SUDAS) for outdoor-to-indoor downlink scenarios. In these works, a multi-antenna AP transmits data using spatial multiplexing in the lower licensed frequency band $f_L$. Concurrently, multiple single-antenna shared UE-side distributed antenna components (SUDACs) equipped with mixers, serving as frequency-translation AF relays, perform frequency conversion to convert the signals to the higher unlicensed frequency bands $f_H$. This enables the single-antenna UE to achieve frequency-division multiplexing. However, the single-antenna UE can only exploit frequency-division multiplexing rather than spatial multiplexing, leading to frequency spectrum inefficiency. Moreover, the direct link between the AP and the UE is not considered and utilized. 

Recently, \cite{ref:UE-CoMIMO} proposes an end-user-centric collaborative MIMO (UE-CoMIMO) to offer diversity, rank, and localization enhancements. With the assistance of a nearby multiple-antenna frequency-translation relay, the multi-antenna UE can exploit both spatial and frequency-division multiplexing. Notably, the direct link between AP and UE is utilized in $f_L$ band for efficient rank augmentation, without the need to occupy additional frequency bands. The system-level simulations show significant performance gains thanks to the device collaboration. However, \cite{ref:UE-CoMIMO} lacks formal problem formulation and analysis in a mathematical manner. Additionally, \cite{ref:UE-CoMIMO} does not jointly consider the direct and relay links, resulting in a lower rate due to the antenna interference. Therefore, in the following sections, we formulate the RACA system in detail and propose the WMMSE and SVD-WF optimizations in centralized and distributed protocols, respectively.

\section{System Model and Problem Formulation}
\label{sec:system_problem}




\subsection{System Model}
\label{ssec:system}
As shown in Fig. \ref{fig:system}(c), we consider an uplink single-user MIMO with $N_u$ antennas at the UE and $N_a$ antennas at the AP, collaborated with a $N_r$-antenna AF relay in proximity. In the RACA system, a full-duplex AF relay can be easily implemented without the self-interference issue \cite{ref:SUDAS2014}, thanks to the separate frequency bands for transmitting and receiving data. It is worth noting that AF relays reduce latency compared to decode-and-forward (DF) relays due to the absence of channel decoding and re-encoding, thus without data buffering at the relay. Consequently, the AF-relay increases the transmission data rate in the RACA system without doubling the latency, which is desirable for latency-critical XR applications.

The UE-relay, UE-AP, and relay-AP channels are denoted as $\Hur\shape{C}{N_r}{N_u}$, $\Hua\shape{C}{N_a}{N_u}$, and $\Hra\shape{C}{N_a}{N_r}$, respectively. We assume quasi-static block-fading channels, where $\Hua, \Hur, \Hra$ remain constant during a certain period known as the channel coherence time. The uplink transmission can be divided into two phases. At the first phase during the $t$-th time slot, the UE transmits $\ve{s}^{f_L}_t\shape{C}{N_s}{1}$ and $\ve{s}^{f_H}_t\shape{C}{N_s}{1}$ to the AP and relay in the $f_L$ and $f_H$ bands, respectively. We assume $\mathbb{E}[\ve{s}^{f_L}_t\ve{s}^{f_LH}_t]=\mathbb{E}[\ve{s}^{f_H}_t\ve{s}^{f_HH}_t]=\ma{I}_{N_s}$. Note that $N_s\leq \min\{N_u, N_a, N_r\}$ represents the number of data streams in each link. The received signals at the relay and AP can be respectively expressed as
\begin{equation}
\ve{y}^{f_H}_{ur,t}=\Hur\Wur\ve{s}^{f_H}_t+\ve{z}_{r,t},
\label{eq_y_ur_phase1}
\end{equation}
\begin{equation}
\ve{y}^{f_L}_{a,t}=\underbrace{\Hua\Wua\ve{s}^{f_L}_t}_\text{From UE}+\underbrace{\ve{y}^{f_L}_{ra,t}}_\text{From relay}+\ve{z}_{a,t}.
\label{eq_y_a_phase1}
\end{equation}

At the second phase during $(t+1)$-th time slot, $\ve{y}^{f_H}_{ur,t}$ is transmitted by the relay, and the signal arrived at the AP is written as 
\begin{equation}
\ve{y}^{f_L}_{ra,t+1}=\Hra\ma{\Psi}\ve{y}^{f_H}_{ur,t}.
\label{eq_y_ra_phase2}
\end{equation}
Therefore, the received signals at the AP can be expressed as
\begin{equation}
\begin{aligned}
\ve{y}^{f_L}_{a,t+1}= &\ \Hua\Wua\ve{s}^{f_L}_{t+1}+\ve{y}^{f_L}_{ra,t+1}+\ve{z}_{a,t+1}\\
= & \ \Hua\Wua\ve{s}^{f_L}_{t+1} + \ve{z}_{a,t+1} \\
& + \Hra\ma{\Psi}(\Hur\Wur\ve{s}^{f_HH}_t+\ve{z}_{r,t}),
\label{eq_y_a_phase2}
\end{aligned}
\end{equation}
where $\Wua\shape{C}{N_u}{N_s}$ and $\Wur\shape{C}{N_u}{N_s}$ are the precoders in the $f_L$ and $f_H$ bands, respectively. The relay amplifying matrix, denoted as $\ma{\Psi}\shape{C}{N_r}{N_r}$, is used to enhance the signal quality and perform the frequency conversion. Let $\ve{z}_{r,t}\shape{C}{N_r}{1}\sim\mathcal{CN}(\ve{0},\sigma_r^2\ma{I}_{N_r})$ and $\ve{z}_{a,t}\shape{C}{N_a}{1}\sim\mathcal{CN}(\ve{0},\sigma_a^2\ma{I}_{N_a})$ be the additive white Gaussian noise (AWGN) at the relay and the AP, and $\sigma_r^2$ and $\sigma_a^2$ are their corresponding noise variance.

After expanding \eqref{eq_y_a_phase2}, the doubled transmit data streams can be concatenated as $\ve{s}_{t+1}=[\ve{s}^{f_LT}_{t+1}, \ve{s}^{f_HT}_{t}]^T\shape{C}{2N_s}{1}$. Then, the corresponding effective channel can be expressed as $\ma{H} = [\Hua\Wua,\Hra\ma{\Psi}\Hur\Wur]\shape{C}{N_a}{2N_s}$. Therefore, \eqref{eq_y_a_phase2} can be rewritten in a compact form as follows
\begin{equation}
\ve{y}^{f_L}_{a,t+1} = \underbrace{\ma{H}\ve{s}_{t+1}}_\text{Desired signal} + \underbrace{(\Hra\ma{\Psi}\ve{z}_{r,t}+\ve{z}_{a,t+1})}_\text{Noise}. 
\label{eq_y_a_phase2_concate}
\end{equation}
 
Then, the achievable rate of the UE can be derived as 
\begin{equation}
R(\Wua, \Wur, \ma{\Psi}) = \text{log}_2|\ma{I}_{2N_s}+\ma{H}^H\ma{J}^{-1}\ma{H}|, 
\label{eq_rate}
\end{equation}
where $\ma{J}=\sigma_r^2(\Hra\ma{\Psi})(\Hra\ma{\Psi})^H+\sigma_a^2\ma{I}_{N_a}\shape{C}{N_a}{N_a}$ is the noise covariance matrix.

\subsection{Channel Model}
\label{ssec:channel}
\begin{figure}[tbp]
\centerline{\includegraphics[width=0.90\linewidth]{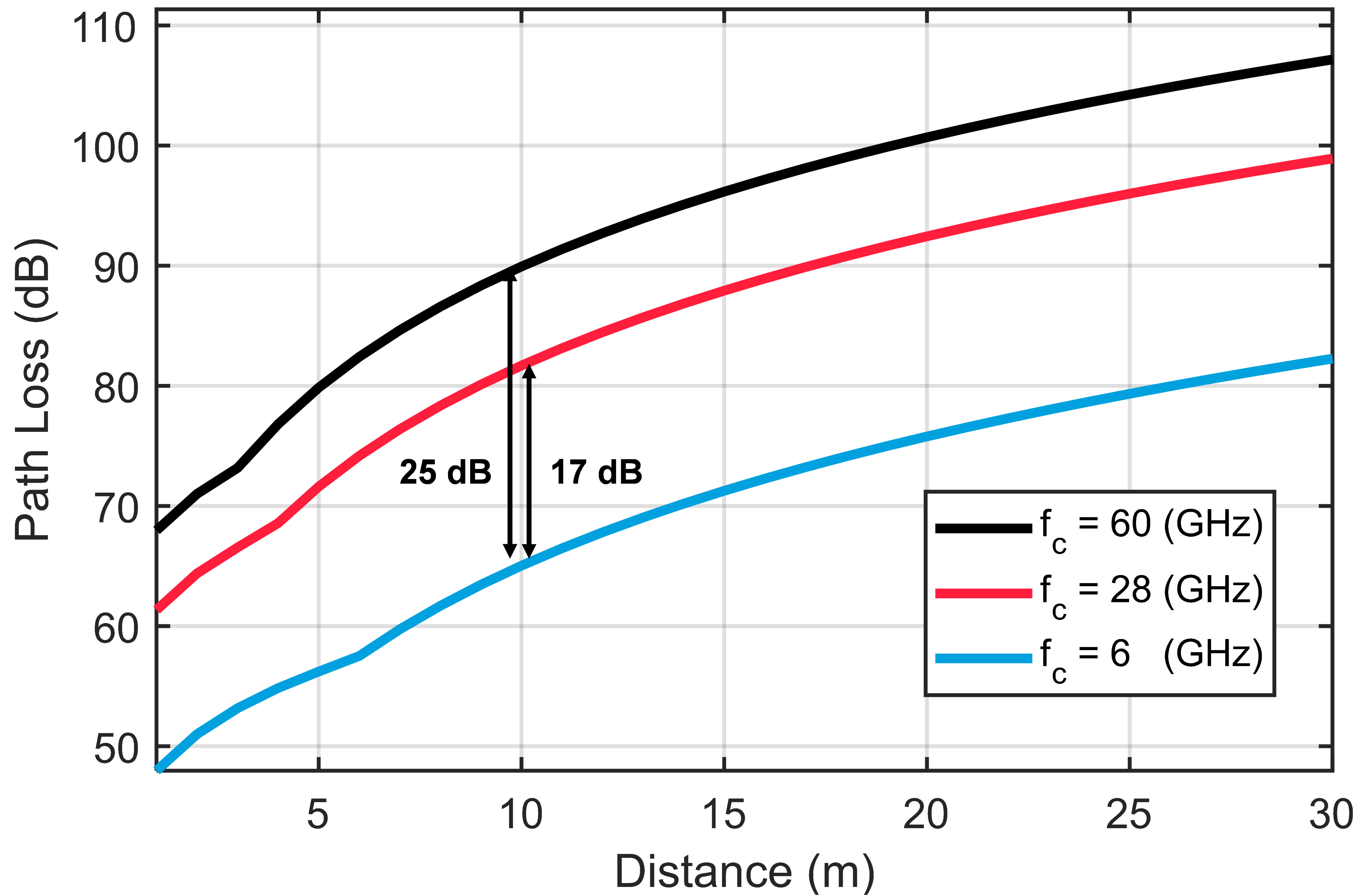}}
\caption{Large-scale path loss of InH scenario.}
\vspace{-1em}
\label{fig:path_loss}
\end{figure}

Although our proposed optimization algorithms can be applied to any channel model, we establish the following channel model to capture realistic indoor environments in XR applications. We adopt the Rayleigh fading channel model in our system. The entries of $\Hua, \Hur, \Hra$ are randomly generated from the independent and identically distribution of $\mathcal{CN}(0,1)$. As for the large-scale fading, to capture the behavior of indoor scenarios for XR usage, we adopt the frequency-dependent Indoor Hotspot (InH) path loss (PL) model included by 3GPP to reflect the different attenuations in $f_L$ and $f_H$ frequency bands. Based on \cite{ref:PL_InH}, the PL in the non-line-of-sight (NLOS) cases can be modeled as
\begin{align}
& PL_\text{InH-NLOS} = \max\{PL_\text{InH-LOS},PL_\text{InH-NLOS}^{'}\},
\label{eq_PL_NLOS} \\
& PL_\text{InH-LOS} \: \: = 32.4 + 17.3\text{log}_{10}(d) + 20\text{log}_{10}(f_c),
\label{eq_PL_LOS} \\
& PL_\text{InH-NLOS}^{'} = 17.3 + 38.3\text{log}_{10}(d) + 24.9\text{log}_{10}(f_c),
\label{eq_eq_PL_NLOS_}
\end{align}
where $d$ represents a transmission distance measured in meters, and $f_c$ denotes a carrier frequency, expressed in gigahertz (GHz). Fig. \ref{fig:path_loss} shows the path loss of different frequencies over the distance. When $f_H$ operates in the higher frequency, e.g., mmWave, the signal may suffer from the additional 17 to 25 dB severe attenuation.

It is worth noting that the three channels can be acquired by the well-known MIMO channel estimation methods, such as least square (LS) or minimum mean square error (MMSE) estimators. Nevertheless, this paper focuses on optimization analysis and assumes the perfect CSI is available.

\subsection{Problem Formulation}
\label{ssec:problem}

Our objective is to maximize the UE's transmission data rate by jointly optimizing the precoders $\Wua, \Wur$, and the relay amplifying matrix $\ma{\Psi}$. The maximization problem can be formulated as
\begin{subequations} \label{P_rate}
\begin{alignat}{2}
& \hspace{-2em} \underset{\Wua, \Wur, \ma{\Psi}}{\text{max}}\ && \text{log}|\ma{I}_{2N_s}+\ma{H}^H\ma{J}^{-1}\ma{H}| \label{P_rate_max} \\
& \text{s.t.} && \lVert\Wua\rVert_F^2\leq \Pua, \label{P_rate_pua} \\
&&& \lVert\Wur\rVert_F^2\leq \Pur, \label{P_rate_pur} \\
&&& \lVert\ma{\Psi}\Hur\Wur\rVert_F^2+\sigma_r^2\lVert\ma{\Psi}\rVert_F^2 \leq P_r. \label{P_rate_pr}
\end{alignat}
\end{subequations} 
In \eqref{P_rate_pua} and \eqref{P_rate_pur}, $\Pua$ and $\Pur$ are the maximal transmit power at the UE in $f_L$ and $f_H$ bands, respectively. Moreover, the power of the transmitted signal from the relay,  i.e., the power of $\ma{\Psi}(\Hur\Wur\ve{s}^{f_H}+\ve{z}_r)$, should satisfy the maximal transmit power $P_r$ at the relay, which is constrained in \eqref{P_rate_pr}.

The coupled variables in the objective \eqref{P_rate_max} and the constraint \eqref{P_rate_pr} make the optimization problem \eqref{P_rate} non-convex and thus intractable. In the following sections, we propose two efficient optimization methods in the centralized and distributed transmission protocols.

\section{Transmission Protocols}
\label{sec:protocol}

\begin{figure*}[tbp]
\centerline{\includegraphics[width=1.0\linewidth]{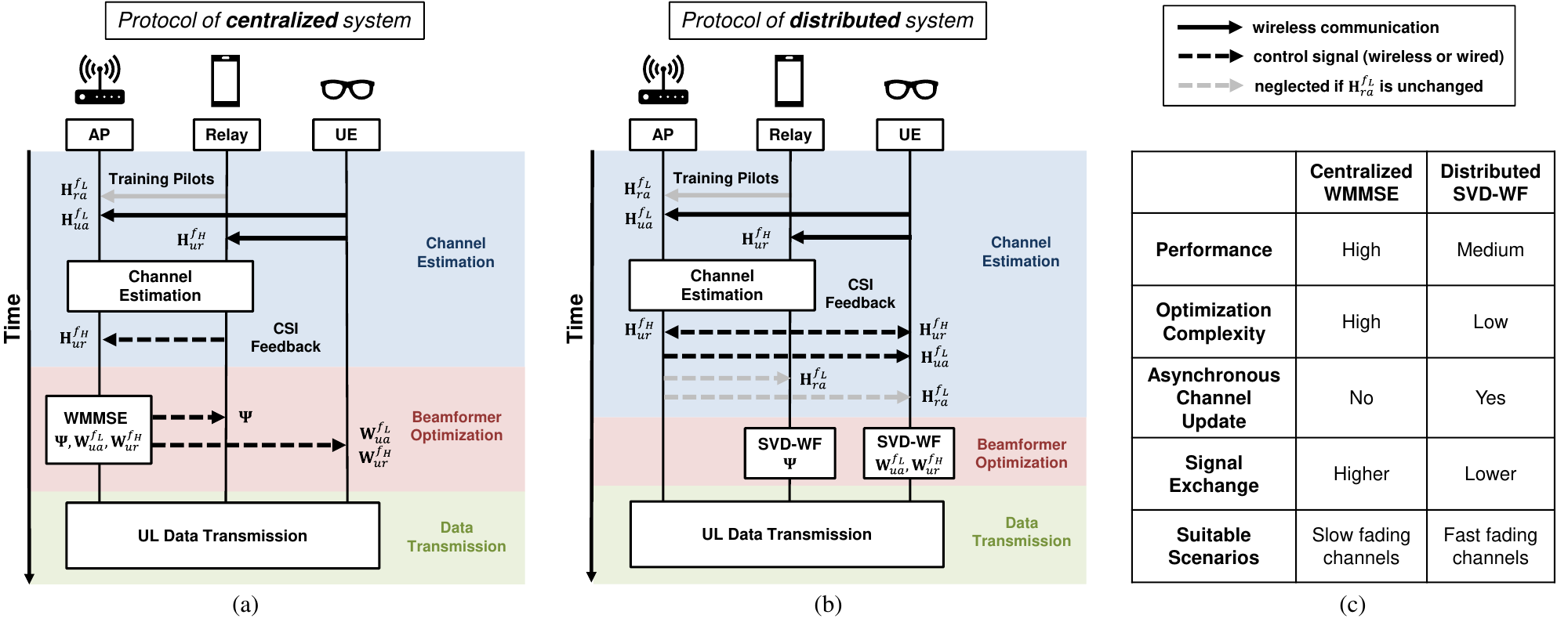}}
\caption{Transmission protocol: (a) the centralized system, (b) the distributed system, and (c) the comparison table.}
\label{fig:protocol}
\end{figure*}

To realize effective device collaboration, it is important to consider both the overhead of channel exchanges and the platform nodes where optimizations are computed. In this paper, we propose centralized and distributed transmission protocols for different scenarios. In a centralized system, an iterative WMMSE-based optimization algorithm is processed at the AP to jointly consider the direct and relay links for antenna interference mitigation. In a distributed system, without considering the antenna interference, the rate maximization problem is heuristically decomposed as two independent sub-problems. The low-complexity SVD-WF algorithms are computed in the relay and UE nodes instead of the AP. The detailed optimization algorithms of WMMSE and SVD-WF will be described in Section \ref{sec:WMMSE} and Section \ref{sec:SVD-WF}, respectively.

As depicted in Fig. \ref{fig:protocol}, at the beginning of both protocols, the UE transmits the pilots to the AP and the relay for $\Hua$ and $\Hur$ estimation, while the relay transmits the pilots to the AP for $\Hra$ estimation. Subsequently, the beamformer variables are optimized and configured before the data transmission. In a centralized system, the AP performs the iterative WMMSE to jointly enhance the UE rate with the $\Hur$ CSI feedback. Once the WMMSE algorithm converges, the optimized variables, i.e., $\Wua, \Wur, \ma{\Psi}$, must be fed back through the control links for deployment. On the other hand, in the distributed system, the low-complexity SVD-WF optimization can be distributively offloaded to the UE and the relay owing to the problem decomposition. Some relevant channels need to be fed back to the UE and relay nodes through the control links. Within a single channel coherence time, the total number of entries to be exchanged via the control link is 
\begin{align}
& \textbf{Centralized system: } N_rN_u+N_r^2+2N_uN_s,
\label{eq_entry_cent} \\
& \textbf{Distributed system: } 2N_rN_u+N_aN_u+2N_aN_r.
\label{eq_entry_dist}
\end{align}

In our simulation setup with $(N_s,N_u,N_r,N_a)=(2,2,4,4)$, the distributed system requires an exchange of 56 entries, while the centralized system only involves 32 entries to be exchanged. Note that it is reasonable to assume that the coherence time of $\Hra$ is longer than those of $\Hur$ and $\Hua$ because the relay, such as a smartphone, may maintain a fixed position relative to the AP in XR usage scenarios \cite{ref:CE_twotimescle}. Thus, a two-timescale channel estimation can be applied to reduce the redundant exchanges, which means the data exchanges in the gray arrows in Fig. \ref{fig:protocol} can be reduced if $\Hra$ is unchanged. However, this asynchronous channel update is only supported in a distributed system. This limitation is inherent to a centralized system because any change in one of the channels necessitates an update of the jointly optimized variables to both the relay and the UE. As a result, a distributed system can avoid $2N_aN_r=32$ entries' feedback in each asynchronous update. Overall, it incurs fewer signal exchanges than a centralized system if the coherence time of $\Hra$ is at least four times longer than those of $\Hur$ and $\Hua$. Note that the low-complexity SVD-WF optimization can still be employed in a centralized protocol if the asynchronous channel update is not exploited.

In summary, the low-complexity SVD-WF in the distributed system is more suitable for fast-fading channels due to its flexibility for asynchronous channel updates. Conversely, the high-performance iterative WMMSE in the centralized system is better suited for slow-fading channels, as it maintains a high data rate without frequent channel updates. The comparisons are summarized in Fig. \ref{fig:protocol}(c).

\section{Centralized Systems and WMMSE-based Solutions}
\label{sec:WMMSE}
In a centralized system, we assume all CSI for each link is known at the AP. To deal with the intractable problem \eqref{P_rate} at the AP, even when we adopt an alternating optimization (AO)-based approach that iteratively optimizes one variable at a time while keeping the others fixed, the problem \eqref{P_rate} with respect to $\ma{\Psi}$ is still non-convex. In this paper, we exploit the equivalence between the rate maximization and the WMMSE according to \cite{ref:WMMSE2008}, \cite{ref:WMMSE2011}. After the problem transformation, the convex single-variable sub-problems can be iteratively solved in an AO-based manner. Once the WMMSE-based optimization converges, the AP feedbacks the optimized variables to the relay and UE for deployment.

\subsection{WMMSE-based Problem Transformation}
\label{ssec:wmmse_problem}

First, a receiver $\Wa\shape{C}{2N_s}{N_a}$ is utilized at the AP to recover the signals as $\tilde{\ve{s}}_{t} = \Wa\ve{y}^{f_L}_{a,t}$, with $N_a \geq 2N_s$ requirement. We define the MSE matrix $\ma{E}\shape{C}{2N_s}{2N_s}$ as 
\begin{align}
    &\quad \quad \quad \quad \quad \ma{E} = \mathbb{E}[(\tilde{\ve{s}_t}-\ve{s}_t)(\tilde{\ve{s}_t}-\ve{s}_t)^H] \notag\\
    &\hspace{-1em}= (\Wa\ma{H}-\ma{I}_{2N_s})(\Wa\ma{H}-\ma{I}_{2N_s})^H + \Wa\ma{J}\Wa^H.
    \label{MSE_matrix}
\end{align}
Then, the equivalent WMMSE problem can be formulated as
\begin{subequations} \label{P_wmmse}
\begin{alignat}{2}
& \hspace{-3em} \underset{\Wa, \Wua, \Wur, \ma{\Psi}, \ma{Z}}{\text{min}}\ && \ \text{tr}(\ma{Z}\ma{E}) -  \text{log}|\ma{Z}| \label{P_wmmse_min} \\
& \ \text{s.t.} && \ \eqref{P_rate_pua}, \eqref{P_rate_pur}, \eqref{P_rate_pr}, \label{P_wmmse_p}
\end{alignat}
\end{subequations} 
where $\ma{Z}=\begin{bmatrix}\ma{Z}_u^T & \ma{Z}_l^T\end{bmatrix}^T\shape{C}{2N_s}{2N_s}\succeq\ma{0}$ is an auxiliary weight matrix, and $\ma{Z}_u, \ma{Z}_l\shape{C}{N_s}{2N_s}$ are the upper and lower submatrices of $\ma{Z}$, respectively. According to \cite{ref:WMMSE2008}, \cite{ref:WMMSE2011}, we have the following theorem to establish the equivalence between problem \eqref{P_rate} and \eqref{P_wmmse}. The detailed proof is in \textit{Appendix A}.

\begin{theorem}
In the RACA system, the WMMSE problem \eqref{P_wmmse} is equivalent to the rate maximization problem \eqref{P_rate}, i.e., the global optimal solutions of both problems are identical.
\end{theorem} \label{theorem_rate_wmmse}

\subsection{Iterative AO-based Optimization}
\label{ssec:wmmse_ao}
Although the problem \eqref{P_wmmse} is convex with respect to each individual variable, its joint optimization remains non-convex. Therefore, we adopt an AO-based approach, iteratively optimizing one variable at a time while keeping the others fixed to obtain single-variable sub-problems. This method ensures that the solution converges to a local optimum because the sub-problems are convex. The proposed WMMSE is summarized in \textbf{Algorithm \ref{alg::wmmse}}, and the five sub-problems are solved in the following details.

\begin{algorithm}[tp]
\caption{Joint WMMSE algorithm}
\label{alg::wmmse}
\begin{algorithmic}[1]
    \INPUT $\Hur$, $\Hua$, $\Hra$, $\sigma_r^2$, $\sigma_a^2$, stopping threshold $\epsilon_{min}^c$, and the maximum number of iteration $T_{max}^c$.
    \OUTPUT Optimized $\Wua$, $\Wur$, $\ma{\Psi}$.
    \Require $t=0$, $\ma{Z}^{(0)} = \ma{I}_{2N_s}$, random initialized $\Wa^{(0)}$, $\Wur^{(0)}$, $\ma{\Psi}^{(0)}$ with power normalization to satisfy \eqref{P_rate_pua}-\eqref{P_rate_pr}, or initialized by \textbf{Algorithm \ref{alg::svd_wf}}.
    \State \textbf{repeat}
    \State \qquad Update $\Wa^{(t+1)}$ by \eqref{Wa_opt}.
    \State \qquad Update $\Wua^{(t+1)}$ by \eqref{Wua_opt}.
    \State \qquad Update $\Wur^{(t+1)}$ by \eqref{Wur_opt}.
    \State \qquad Update $\ma{\Psi}^{(t+1)}$ by \eqref{Psi_opt}.
    \State \qquad Update $\ma{Z}^{(t+1)}$ by \eqref{Z_opt}.
    \State \qquad $\epsilon^{(t+1)} = (R^{(t+1)} - R^{(t)})/R^{(t+1)}$ with \eqref{eq_rate}. 
    \State \qquad $t = t + 1$.
    \State \textbf{until} $\epsilon^{(t)} < \epsilon_{min}^c$ or $t > T_{max}^c$
\end{algorithmic}
\end{algorithm}

\subsubsection{Optimization of $\Wa$}
Firstly, the problem \eqref{P_wmmse} with respect to $\Wa$ becomes an unconstrained convex problem, thus the closed-form optimal solution can be derived as 
\begin{equation}
    \Wa = \ma{H}^H(\ma{H}\ma{H}^H+\ma{J})^{-1},
    \label{Wa_opt}
\end{equation}
which is a well-known MMSE receiver to recover the signals with minimal MSE.
\subsubsection{Optimization of $\Wua$}
With respect to $\Wua$, the formulation in \eqref{P_wmmse} can be expressed as
\begin{subequations} \label{P_wmmse_Wua}
\begin{alignat}{2}
& \underset{\Wua}{\text{min}}\ && \ \text{tr}(\Wua^H\ua{B}\Wua) - 2 \text{Re}\{ \text{tr}(\Wua^H\ua{C}^H)\} \label{P_wmmse_Wua_min} \\ 
& \ \text{s.t.} && \ \text{tr}(\Wua^H\Wua) \leq \Pua, \label{P_wmmse_Wua_pua}
\end{alignat}
\end{subequations}
where $\ua{B} = \Hua^H\Wa^H\ma{Z}\Wa\Hua\shape{C}{N_u}{N_u}$, and $\ua{C}=\ma{Z}_{u}\Wa\Hua\shape{C}{N_s}{N_u}$. Based on Karush-Kuhn-Tucker (KKT) conditions \cite{ref:boyd2004convex}, the optimal $\Wua$ is
\begin{equation}
    \Wua = (\ua{B}+\nu_{ua}^{f_L}\ma{I}_{N_u})^{-1}\ua{C}^H,
    \label{Wua_opt}
\end{equation}
where $\nu_{ua}^{f_L}$ is the Lagrange multiplier. Based on the complementary slackness condition \cite{ref:boyd2004convex}, if the constraint \eqref{P_wmmse_Wua_pua} is not satisfied when $\nu_{ua}^{f_L} = 0$,  the optimal $\nu_{ua}^{f_L*}$ will be determined through a bisection search to make the equality holds in \eqref{P_wmmse_Wua_pua}. 

Furthermore, to avoid matrix inversions in the bisection search, we exploit the positive semidefinite property of $\ua{B}$ to achieve matrix inversion-free. Specifically, since $\ua{B}\succeq\ma{0}$, we first let $\ua{B}=\ua{U_B}\ua{\Lambda_B}\ua{U_B}^H$ by SVD. Then, $\Wua=\ua{U_B}(\ua{\Lambda_B}+\nu_{ua}^{f_L}\ma{I}_{N_u})^{-1}\ua{U_B}^H\ua{C}^H$. Let $\ua{X}=\ua{U_B}^H\ua{C}^H\ua{C}\ua{U_B}\shape{C}{N_u}{N_u}$
Then, we have
\begin{align}
    \text{tr}(\Wua^H\Wua) & = \text{tr}((\ua{\Lambda_B}+\nu_{ua}^{f_L}\ma{I}_{N_u})^{-2}\ua{X}) \notag\\
    & = \sum_{i=1}^{N_u} \frac{[\ua{X}]_{i,i}}{([\ua{\Lambda_B}]_{i,i}+\nu_{ua}^{f_L})^2}.
    \label{nu_Wua_diag}
\end{align}
Note that the transmit power expressed in \eqref{nu_Wua_diag} monotonically decreases with the increased $\nu_{ua}^{f_L}$ to meet the power constraint $\Pua$. Moreover, according to the following inequality
\begin{equation}
   \sum_{i=1}^{N_u} \frac{[\ua{X}]_{i,i}}{(\nu_{ua}^{f_L*})^2} \geq \sum_{i=1}^{N_u} \frac{[\ua{X}]_{i,i}}{([\ua{\Lambda_B}]_{i,i}+\nu_{ua}^{f_L*})^2} = \Pua,
\end{equation}
we can derive that $\nu_{ua}^{f_L*}\leq\sqrt{\frac{\text{tr}(\ua{X})}{\Pua}}$. With the closed-form upper bound, the bisection search space can be restricted.
\subsubsection{Optimization of $\Wur$}
With respect to $\Wur$, the sub-problem of \eqref{P_wmmse} is represented as
\begin{subequations} \label{P_wmmse_Wur}
\begin{alignat}{2}
& \hspace{-0.3em}\underset{\Wur}{\text{min}} && \ \text{tr}(\Wur^H\ur{B}\Wur) - 2 \text{Re}\{ \text{tr}(\Wur^H\ur{C}^H)\} \label{P_wmmse_Wur_min} \\ 
&  \text{s.t.} && \ \text{tr}(\Wur^H\Wur) \leq \Pur, \label{P_wmmse_Wur_pua} \\
&&& \ \text{tr}(\Wur^H\ur{\Tilde{B}}\Wur) \leq P_r - \sigma_r^2\lVert\ma{\Psi}\rVert_F^2, \label{P_wmmse_Wur_pr}
\end{alignat}
\end{subequations}
where $\ur{B} = \Hur^H\ma{\Psi}^H\Hra^H\Wa^H\ma{Z}\Wa\Hra\ma{\Psi}\Hur\shape{C}{N_u}{N_u}$, $\ur{C}=\ma{Z}_{l}\Wa\Hra\ma{\Psi}\Hur\shape{C}{N_s}{N_u}$, and $\ur{\Tilde{B}}=\Hur^H\ma{\Psi}^H\ma{\Psi}\Hur\shape{C}{N_u}{N_u}$. Thus, the optimal $\Wur$ is expressed as
\begin{equation}
    \Wur = (\ur{B}+\nu_{ur}^{f_H}\ma{I}_{N_u}+\tilde{\nu}_{r}\ur{\Tilde{B}})^{-1}\ur{C}^H,
    \label{Wur_opt}
\end{equation}
where the Lagrange multiplier $\nu_{ur}^{f_H}$ and $\tilde{\nu}_{r}$ can only be obtained by two-layer bisection search \cite{ref:relay_hybrid} to satisfy \eqref{P_wmmse_Wur_pua} and \eqref{P_wmmse_Wur_pr}.
\subsubsection{Optimization of $\ma{\Psi}$}
With respect to $\ma{\Psi}$, the sub-problem of \eqref{P_wmmse} is formulated as
\begin{subequations} \label{P_wmmse_Psi}
\begin{alignat}{2}
& \underset{\ma{\Psi}}{\text{min}}\ && \ \text{tr}(\underPsi{D}^H\ma{\Psi}^H\underPsi{B}\ma{\Psi}\underPsi{D}) - 2 \text{Re}\{ \text{tr}(\ma{\Psi}^H\underPsi{C}^H)\} \notag\\
&&& \ +\text{tr}(\ma{\Psi}^H\sigma_r^2\underPsi{B}\ma{\Psi}) \label{P_wmmse_Psi_min}\\ 
& \ \text{s.t.} && \ \text{tr}(\ma{\Psi}(\underPsi{D}\underPsi{D}^H+\sigma_r^2\ma{I}_{N_r})\ma{\Psi}^H) \leq P_r, \label{P_wmmse_Psi_pr}
\end{alignat}
\end{subequations}
where $\underPsi{B}=\Hra^H\Wa^H\ma{Z}\Wa\Hra\shape{C}{N_r}{N_r}$, $\underPsi{C}=\Hur\Wur\ma{Z}_l\Wa\Hra\shape{C}{N_r}{N_r}$, and $\underPsi{D}=\Hur\Wur\shape{C}{N_r}{N_s}$. Therefore, the optimal $\ma{\Psi}$ is given as
\begin{equation}
    \ma{\Psi} = (\underPsi{B}+\nu_{\Psi}\ma{I}_{N_r})^{-1}\underPsi{C}^H(\underPsi{D}\underPsi{D}^H+\sigma_r^2\ma{I}_{N_r})^{-1},
    \label{Psi_opt}
\end{equation}
where the Lagrange multiplier $\nu_{\Psi}$ can be chosen via bisection search with matrix inversion-free diagonal structure as \eqref{nu_Wua_diag}.
\subsubsection{Optimization of $\ma{Z}$}
Lastly, due to the convexity without constraints, the optimal $\ma{Z}$ can be derived by setting the derivative of $\eqref{P_wmmse_min}$ as zero, which implies
\begin{equation}
    \ma{Z} = (\ma{E})^{-1}.
    \label{Z_opt}
\end{equation}

\section{Distributed Systems and SVD-WF Solutions}
\label{sec:SVD-WF}
In this section, we introduce a low-complexity algorithm of SVD with water-filling, aiming for an optimization distributively processed at the UE and relay nodes in the distributed protocol system. Owing to the limited exchanges of CSI at each node, ensuring a system performance that reaches a local optimum becomes inherently challenging. Consequently, we heuristically decompose the problem \eqref{P_rate} into two independent sub-problems, each aimed at maximizing the data rates of the direct or relay link. While this approach does not explicitly mitigate antenna interference at AP between two links, the simulation results will demonstrate its effectiveness. The detailed steps of the proposed SVD-WF are summarized in \textbf{Algorithm \ref{alg::svd_wf}}.

\begin{algorithm}[tp]
\caption{Separate SVD-WF algorithm}
\label{alg::svd_wf}
\begin{algorithmic}[1]
    \INPUT $\Hur$, $\Hua$, $\Hra$, $\sigma_r^2$, $\sigma_a^2$, stopping threshold $\epsilon_{min}^d$, and the maximum number of iteration $T_{max}^d$. 
    \OUTPUT Optimized $\Wua$, $\Wur$, $\ma{\Psi}$.
    \Require $t=0$.
    \State Perform SVD on the channels as \eqref{SVD_Hua}, \eqref{SVD_Hur}, \eqref{SVD_Hra}.
    \\
    \State $\slash\slash$ $\textit{Optimization of direct link}$
    \State Obtain $\Wua$ by \eqref{Wua_optimal}.
    \\
    \State $\slash\slash$ $\textit{Optimization of relay link}$
    \State \textbf{repeat}
    \State \qquad Update $\tilde{\ve{p}}_{u}^{f_H(t+1)}$ by \eqref{p_tilde_opt}, given fixed $\tilde{\ve{p}}_{\ma{\Psi}}^{(t)}$.
    \State \qquad Update $\tilde{\ve{p}}_{\ma{\Psi}}^{(t+1)}$ by \eqref{p_tilde_opt}, given fixed $\tilde{\ve{p}}_{u}^{f_H(t)}$.
    \State \qquad Calculate $\epsilon_1^{(t+1)} = \lVert \tilde{\ve{p}}_{u}^{f_H(t+1)} - \tilde{\ve{p}}_{u}^{f_H(t)}\rVert_1$. 
    \State \qquad Calculate $\epsilon_2^{(t+1)} = \lVert \tilde{\ve{p}}_{\ma{\Psi}}^{(t+1)} - \tilde{\ve{p}}_{\ma{\Psi}}^{(t)}\rVert_1$.
    \State \qquad $t = t + 1$.
    \State \textbf{until} ($\epsilon_1^{(t)} < \epsilon_{min}^d$ and $\epsilon_2^{(t)} < \epsilon_{min}^d$) or $t > T_{max}^d$
    \State Let $p_{u,i}^{f_H} = \tilde{p}_{u,i}^{f_H}$ and $p_{\ma{\Psi},i} = \tilde{p}_{\ma{\Psi},i}/(\lambda_{ur,i}^{f_H2}p_{u,i}^{f_H}+\sigma_r^2), \forall i$.
    \State Obtain $\Wur$ and $\ma{\Psi}$ by \eqref{Wur_optimal} and \eqref{Psi_optimal}, respectively.
    
\end{algorithmic}
\end{algorithm}

\subsection{SVD Optimization of the Direct Link}
\label{ssec:SVD-direct}
The decomposed rate maximization of the direct link is essentially a point-to-point MIMO system, which can be expressed as
\begin{subequations} \label{P_rate_direct}
\begin{alignat}{2}
& \underset{\Wua}{\text{max}}\ && \quad \text{log}|\ma{I}_{N_s}+\frac{1}{\sigma_a^2}\Wua^H\Hua^H\Hua\Wua| \label{P_rate_direct_max} \\
& \ \text{s.t.} && \quad \lVert\Wua\rVert_F^2\leq \Pua. \label{P_rate_direct_pua}
\end{alignat}
\end{subequations} 
The well-known optimal SVD-WF solution aims to diagonalize the channel $\Hua$, resulting in parallel sub-channel beamforming \cite{ref:diag_mimo} . After the SVD process, the channel $\Hua$ can be decomposed as
\begin{equation}
    \Hua \ = \ \ua{U}\ua{\Lambda}\ua{V}^H,
\label{SVD_Hua}
\end{equation}
where the unitary $\ua{U}\shape{C}{N_a}{N_a}$ and $\ua{V}\shape{C}{N_u}{N_u}$ are the left and right singular vectors, respectively. $\ua{\Lambda}\shape{C}{N_a}{N_u}$ is a rectangular diagonal matrix whose diagonal elements are the singular values in descending order. Then, the diagonal vector of $\ua{\Lambda}$ can be written as $\text{diag}(\ua{\Lambda})=[\lambda_{ua,1}^{f_L}, ..., \lambda_{ua,N_s}^{f_L}, ... ,\lambda_{ua,N_{ua}^{min}}^{f_L}]^T\shape{R}{N_{ua}^{min}}{1}_+$, where $N_{ua}^{min}=\min\{N_u, N_a\}$. Then, the optimal precoder can be given as 
\begin{equation}
    \Wua = \ua{\Tilde{V}}\ma{\Lambda}_u^{f_L},
\label{Wua_optimal}
\end{equation}
where $\ua{\Tilde{V}}\shape{C}{N_u}{N_s}$ is the $N_s$ leftmost columns of $\ua{V}$, and $\ma{\Lambda}_u^{f_L}\shape{R}{N_s}{N_s}_+$ is a diagonal matrix to determine the power allocation at each sub-channel. We denote $\text{diag}(\ma{\Lambda}_u^{f_L}) = [\sqrt{p_{u,1}^{f_L}}, ... , \sqrt{p_{u,N_s}^{f_L}}]^T\shape{R}{N_s}{1}_+$. Therefore, the scalarized problem can be formulated as
\begin{subequations} \label{P_rate_direct_scale}
\begin{alignat}{2}
& \underset{\{p_{u,i}^{f_L}\}_{i=1}^{N_s}}{\text{max}}\ && \quad \sum_{i=1}^{N_s}\text{log}(1+\gamma_{ua,i}^{f_L} p_{u,i}^{f_L}) \label{P_rate_direct_scale_max} \\
& \quad \ \text{s.t.} && \quad \sum_{i=1}^{N_s}p_{u,i}^{f_L} \leq \Pua, \label{P_rate_direct_scale_pua}
\end{alignat}
\end{subequations}
where $\gamma_{ua,i}^{f_L} = \lambda_{ua,i}^{f_L2}/\sigma_a^2$ is the equivalent channel-to-noise ratio (CNR) on the i-th sub-channel. Consequently, the optimal power allocation can be derived as $p_{u,i}^{f_L} = (1/\nu_d - 1/\gamma_{ua,i}^{f_L})_+$, where $\nu_d$ is the Lagrange multiplier to satisfy constraint \eqref{P_rate_direct_scale_pua}.


\subsection{SVD Optimization of the Relay Link}
\label{ssec:SVD-relay}
The decomposed rate maximization of the relay link can regarded as the two-hop MIMO system, which is given by
\begin{subequations} \label{P_rate_relay}
\begin{alignat}{2}
& \underset{\Wur, \ma{\Psi}}{\text{max}}\ && \quad \text{log}|\ma{I}_{N_s}+\ma{\tilde{H}}^H\ma{J}^{-1}\ma{\tilde{H}}| \label{P_rate_relay_max} \\
& \quad \text{s.t.} && \quad \eqref{P_rate_pur}, \eqref{P_rate_pr}, \label{P_rate_relay_pur_pr}
\end{alignat}
\end{subequations} 
where $\ma{\tilde{H}}=\Hra\ma{\Psi}\Hur\Wur$. As the point-to-point MIMO system, \cite{ref:diag_relay} extends the concept to a two-hop relay system and proves that the optimal precoder and relay amplifying matrix will jointly diagonalize the source-relay-destination channel. Please refer to [\citenum{ref:diag_relay}, $\emph{Theorem 1}$] for detailed theorem and proof. Then, we decompose the channel $\Hur$ and $\Hra$ by SVD as
\begin{equation}
    \Hur \ = \ \ur{U}\ur{\Lambda}\ur{V}^H,
\label{SVD_Hur}
\end{equation}
\begin{equation}
    \Hra \ = \ \ra{U}\ra{\Lambda}\ra{V}^H.
\label{SVD_Hra}
\end{equation}
As previously defined, $\ur{U}\shape{C}{N_r}{N_r}$, $\ra{U}\shape{C}{N_a}{N_a}$, $\ur{V}\shape{C}{N_u}{N_u}$, and $\ra{V}\shape{C}{N_r}{N_r}$ are the unitary singular vectors. Then, the singular values are represented as $\text{diag}(\ur{\Lambda})=[\lambda_{ur,1}^{f_H}, ..., \lambda_{ur,N_s}^{f_H}, ... ,\lambda_{ur,N_{ur}^{min}}^{f_H}]^T\shape{R}{N_{ur}^{min}}{1}_+$, $\text{diag}(\ra{\Lambda})=[\lambda_{ra,1}^{f_L}, ..., \lambda_{ra,N_s}^{f_L}, ... ,\lambda_{ra,N_{ra}^{min}}^{f_L}]^T\shape{R}{N_{ra}^{min}}{1}_+$, where $N_{ur}^{min}=\min\{N_u, N_r\}$ and $N_{ra}^{min}=\min\{N_r, N_a\}$. 

The optimal precoder and relay amplifying matrix can be derived as 
\begin{align}
    \Wur &= \ur{\Tilde{V}}\ma{\Lambda}_u^{f_H}, \label{Wur_optimal}\\ 
    \ma{\Psi} \quad &= \ra{\Tilde{V}}\ma{\Lambda}_{\ma{\Psi}}\ur{\Tilde{U}}^H. \label{Psi_optimal}
\end{align}
$\ur{\Tilde{V}}\shape{C}{N_u}{N_s}$, $\ra{\Tilde{V}}\shape{C}{N_r}{N_s}$, $\ur{\Tilde{U}}\shape{C}{N_r}{N_s}$ are the $N_s$ leftmost columns of $\ur{V}$, $\ra{V}$, and $\ur{U}$, respectively. For the power allocation at each sub-channel, we let $\text{diag}(\ma{\Lambda}_u^{f_H}) = [\sqrt{p_{u,1}^{f_H}}, ... , \sqrt{p_{u,N_s}^{f_H}}]^T\shape{R}{N_s}{1}_+$, and $\text{diag}(\ma{\Lambda}_{\ma{\Psi}}) = [\sqrt{\vphantom{|} p_{\ma{\Psi},1}}, ... , \sqrt{\vphantom{|} p_{\ma{\Psi},N_s}}]^T\shape{R}{N_s}{1}_+$. As a result, the problem \eqref{P_rate_relay} can also be diagonalized as multiple single-input single-output (SISO) sub-channels in the following scalar form
\begin{subequations} \label{P_rate_relay_scale}
\begin{alignat}{2}
&\hspace{-1.5em} \underset{\{p_{u,i}^{f_H}, \ p_{\ma{\Psi},i}\}_{i=1}^{N_s}}{\text{max}}\ && \ \sum_{i=1}^{N_s}\text{log}(1+\frac{\lambda_{ra,i}^{f_L2}p_{\ma{\Psi},i}\lambda_{ur,i}^{f_H2}p_{u,i}^{f_H}}{\sigma_r^2\lambda_{ra,i}^{f_L2}p_{\ma{\Psi},i}+\sigma_a^2}) \label{P_rate_relay_scale_max} \\
& \ \ \text{s.t.} && \ \sum_{i=1}^{N_s}p_{u,i}^{f_H} \leq \Pur, \label{P_rate_relay_scale_pur} \\
&&& \ \sum_{i=1}^{N_s} p_{\ma{\Psi},i}(\lambda_{ur,i}^{f_H2}p_{u,i}^{f_H}+\sigma_r^2) \leq P_r. \label{P_rate_relay_scale_pr}
\end{alignat}
\end{subequations}
Next, we define $\tilde{p}_{u,i}^{f_H} = p_{u,i}^{f_H}$ and $\tilde{p}_{\ma{\Psi},i}=p_{\ma{\Psi},i}(\lambda_{ur,i}^{f_H2}p_{u,i}^{f_H}+\sigma_r^2)$ for substitution. Hence, problem \eqref{P_rate_relay_scale} can further simplified as
\begin{subequations} \label{P_rate_relay_scale_tilde}
\begin{alignat}{2}
&\hspace{-0.8em} \underset{\{\tilde{p}_{u,i}^{f_H}, \ \tilde{p}_{\ma{\Psi},i}\}_{i=1}^{N_s}}{\text{max}}\ && \sum_{i=1}^{N_s}\text{log}(1+\frac{\gamma_{ur,i}^{f_H}\tilde{p}_{u,i}^{f_H}\gamma_{ra,i}^{f_L}\tilde{p}_{\ma{\Psi},i}}{1+\gamma_{ur,i}^{f_H}\tilde{p}_{u,i}^{f_H}+\gamma_{ra,i}^{f_L}\tilde{p}_{\ma{\Psi},i}}) \label{P_rate_relay_scale_tilde_max} \\
& \ \quad \text{s.t.} && \sum_{i=1}^{N_s} \tilde{p}_{u,i}^{f_H} \leq \Pur, \label{P_rate_relay_scale_tilde_pur} \\
&&& \sum_{i=1}^{N_s} \tilde{p}_{\ma{\Psi},i} \leq P_r, \label{P_rate_relay_tilde_scale_pr}
\end{alignat}
\end{subequations}
where $\gamma_{ur,i}^{f_H}=\lambda_{ur,i}^{f_H2}/\sigma_r^2$ and $\gamma_{ra,i}^{f_L}=\lambda_{ra,i}^{f_L2}/\sigma_a^2$ are the equivalent CNR on the $i$-th sub-channel.

The non-convex problem \eqref{P_rate_relay_scale_tilde} can be optimally solved by the two-dimensional grid search algorithm in \cite{ref:relay_grid}. However, \cite{ref:diag_relay} shows that the guarantee of global optimality highly depends on the grid density. In addition, the complexity of grid search increases quadratically with $N_s$ and the number of grid points, which is complexity-prohibited in practical deployment. Therefore, we exploit an efficient AO-based algorithm proposed in \cite{ref:relay_ao} to allocate the power alternatively at the UE and the relay. It can be observed that the problem \eqref{P_rate_relay_scale_tilde} can be decomposed into two separate sub-problems in the same mathematical structure with respect to $\{\tilde{p}_{u,i}^{f_H}\}_{i=1}^{N_s}$ or $\{\tilde{p}_{\ma{\Psi},i}\}_{i=1}^{N_s}$. Therefore, the sub-problems can be generally formulated as
\begin{subequations} \label{P_rate_relay_subproblem}
\begin{alignat}{2}
&\underset{\{\tilde{p}_i\}_{i=1}^{N_s}}{\text{max}}\ && \quad \sum_{i=1}^{N_s}\text{log}(1+\frac{a_i b_i\tilde{p}_i}{1+a_i+b_i\tilde{p}_i}) \label{P_rate_relay_subproblem_max} \\
& \ \text{s.t.} && \quad \sum_{i=1}^{N_s} \tilde{p}_i \leq P, \label{P_rate_relay_subproblem_p}
\end{alignat}
\end{subequations}
where $(a_i,b_i,P,\tilde{p}_i)$
\begin{align}
    \hspace{-1em}= &
    \begin{cases}
        (\gamma_{ra,i}^{f_L}\tilde{p}_{\ma{\Psi},i}, \gamma_{ur,i}^{f_H}, P_{ur}^{f_H}, \tilde{p}_{u,i}^{f_H}), & \hspace{-0.5em}\text{with fixed } \{\tilde{p}_{\ma{\Psi},i}\}_{i=1}^{N_s} \\
        (\gamma_{ur,i}^{f_H}\tilde{p}_{u,i}^{f_H}, \gamma_{ra,i}^{f_L}, P_r, \tilde{p}_{\ma{\Psi},i}), & \hspace{-0.5em}\text{with fixed } \{\tilde{p}_{u,i}^{f_H}\}_{i=1}^{N_s} \notag\\
    \end{cases}
    \\ & \quad \forall i = 1, ..., N_s.
\end{align}
Owing to the convexity of the problem \eqref{P_rate_relay_subproblem}, the KKT condition can be employed to derive the optimal solution as
\begin{equation}
    \tilde{p}_i = \frac{1}{b_i}\left[\frac{a_i}{2}\left(\sqrt{1+\frac{4b_i}{a_i\nu_r}}-1\right)-1\right]_+, \ \forall i,
\label{p_tilde_opt}
\end{equation}
where $\nu_r$ is the Lagrange multiplier chosen to satisfy the constraint \eqref{P_rate_relay_subproblem_p}.

After several alternative iterations, the converged solution is guaranteed to be a local optimum due to the monotonically non-decreasing improvement in two separate sub-problems.

\section{Simulation Results}
In this section, we first setup system parameters for indoor XR application scenarios. Then, we introduce and define several benchmarks to ensure a fair comparative analysis. Finally, we evaluate the performance of the RACA system over different aspects, demonstrating its efficiency and offering insightful analysis.

\subsection{Simulation Settings}
\label{ssec:setting}
In the indoor XR scenarios, it is reasonable to assume that the relay, such as a smartphone, is in close proximity to the UE, such as XR glasses, as the user may utilize their own high-capability smartphone to enhance the XR experience. The absolute locations of the UE, relay, and AP are illustrated in Fig. \ref{fig:system}(c) where the distance between the UE and the relay is 1 meter, and their perpendicular distance to the AP is 10 meters. The resource-limited UE is only equipped with $N_u=2$ antennas, assisted by a more powerful relay composed of $N_r=4$ antennas. The AP comprises $N_a=4$ antennas to receive signals. The UE transmits two distinct data streams, with $N_s=2$, at frequency $f_L=6$ GHz and $f_H=28$ GHz. The transmit powers of different frequency bands at the UE are set as $\Pua=\Pur=10$ dBm, and the relay's transmit power is $P_r=10$ dBm. The noise variances are set as $\sigma_r^2=\sigma_a^2=\sigma^2=-90$ dBm. All simulations are averaged over 1000 channel realizations. With respect to the proposed iterative algorithms, we set $T_{max}^c=10^4$, $T_{max}^d=10^2$, and $\epsilon_{min}^c=\epsilon_{min}^d=10^{-7}$ as stopping criteria.

\subsection{Benchmarks}
\label{ssec:benchmark}
The benchmarks' naming principle is structured to begin with the system, followed by the associated optimization algorithm. Firstly, our proposed algorithms are
\begin{itemize}[leftmargin=*]
    \item \AlgoName{RACA-WMMSE}: In the RACA system, the rate is maximized by the centralized WMMSE optimization, as proposed in Section \ref{sec:WMMSE}.
    \item \AlgoName{RACA-SVD(-WF)}: In the RACA system, the rate is maximized by the distributed SVD optimization without (or with) water-filling, as proposed in Section \ref{sec:SVD-WF}.
\end{itemize}
To demonstrate the efficiency of our RACA system, we also include the following three benchmarks. For a fair comparison, different systems have the same total transmit power. 
\begin{itemize}[leftmargin=*]
    \item \AlgoName{CA-SVD-WF}: In the conventional CA system, the UE directly transmits $2N_s$ data streams to the AP through $\ma{H}_{ua}^{f_L}$ and $\ma{H}_{ua}^{f_H}$, with the transmit powers of $\Pua$ and $P_{ua}^{f_H}=\Pur + P_r$, respectively. The CA system can be regarded as two separate point-to-point MIMO systems operating on different frequencies, which are optimally solved by the SVD-WF proposed in Section \ref{ssec:SVD-direct}.
    \item \AlgoName{RA-WMMSE}: In the conventional RA system, the UE simultaneously transmits $N_s$ signals with $P_{u}^{f_L}=\Pua+\Pur$ power to the AP and the relay in the same $f_L$ frequency. Then, the relay forwards these $N_s$ signals to the AP using $P_r$ power, also in the $f_L$ frequency. The received signals at the AP are
    \begin{equation}
    \hat{\ve{s}}^{f_L} = \hat{\ma{W}}_a^{f_L} \hat{\ma{H}}\ve{s}^{f_L} + \hat{\ma{W}}_a^{f_L}(\Hra\ma{\Psi}\ve{z}_r+\ve{z}_a), 
    \label{eq_s_hat_concate_ra}
    \end{equation}
    where $\hat{\ma{H}}=(\Hra\ma{\Psi}\ma{H}_{ur}^{f_L}+\Hua)\hat{\ma{W}}_u^{f_L}\shape{C}{N_a}{N_s}$ is the effective channel, $\ma{H}_{ur}^{f_L}$ is the UE-relay channel in $f_L$ frequency, $\hat{\ma{W}}_u^{f_L}\shape{C}{N_u}{N_s}$ is the precoder, and $\hat{\ma{W}}_a^{f_L}\shape{C}{N_s}{N_a}$ is the receiver. The formulation is similar to the RACA system, and the rate can be iteratively maximized by the WMMSE proposed in Section \ref{sec:WMMSE}.
    
    \item \AlgoName{MIMO-SVD-WF}: As a conventional point-to-point MIMO, the UE directly transmits $N_s$ signals to the AP via $\ma{H}_{ua}^{f_L}$ only, transmitting at a power of $P_{ua}=\Pua+\Pur+P_r$. The rate can be optimally maximized by SVD-WF proposed in Section \ref{ssec:SVD-direct}.
\end{itemize}



\begin{figure}[tbp]
\centerline{\includegraphics[width=0.85\linewidth]{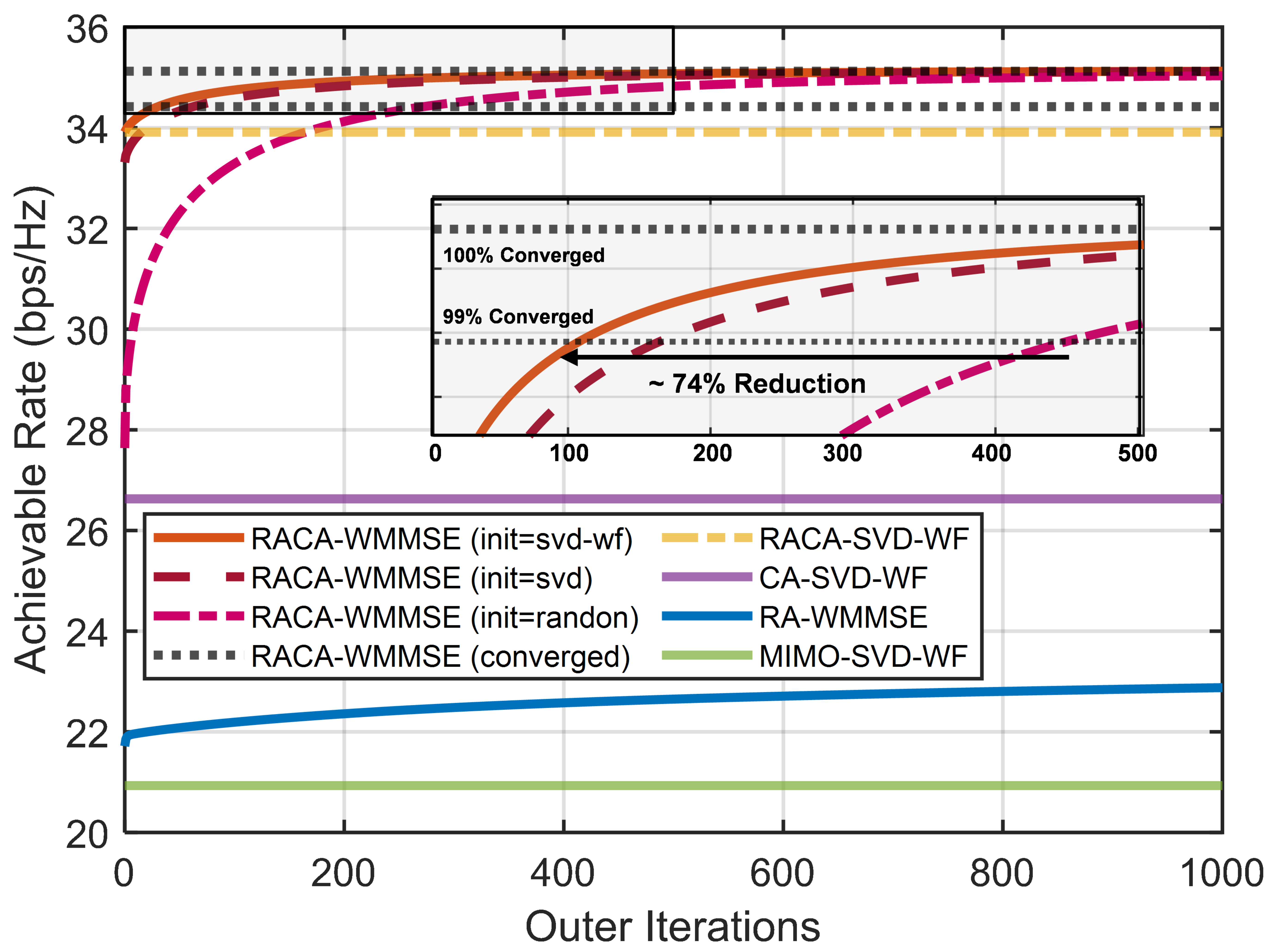}}
\caption{Convergence behavior of the WMMSE.}
\label{fig:convergence}
\end{figure}

\subsection{WMMSE Convergence with Different Initializations}
\label{ssec:convergence_rate}
Since the rate maximization problem is non-convex, the converged local optimal points of iterative WMMSE usually highly depend on the initial points. Therefore, as shown in Fig. \ref{fig:convergence}, we focus on different initializations of WMMSE for convergence speed analysis. Note that the converged \AlgoName{RACA-WMMSE} outperforms non-iterative \AlgoName{RACA-SVD-WF} because it jointly maximizes the rates of two links to mitigate the antenna interference at the AP. Firstly, the random initialization without problem structure starts at a worse rate and converges slowly. Therefore, we exploit the better structural SVD(-WF) initialization for faster convergence. Since the SVD-WF considers the different strengths of sub-channels for allocation, this initialization provides better structure for WMMSE convergence, thus converging slightly faster than the SVD initialization with equal allocation. Consequently, the SVD-WF initialization can reduce 74$\%$ and 89$\%$ iterations to reach 99$\%$ and 98$\%$ converged performance, respectively, compared to the random initialization. As a result, the following \AlgoName{RACA-WMMSE} uses \AlgoName{RACA-SVD-WF} as its initialization.

\begin{figure}[tbp]
\centerline{\includegraphics[width=0.85\linewidth]{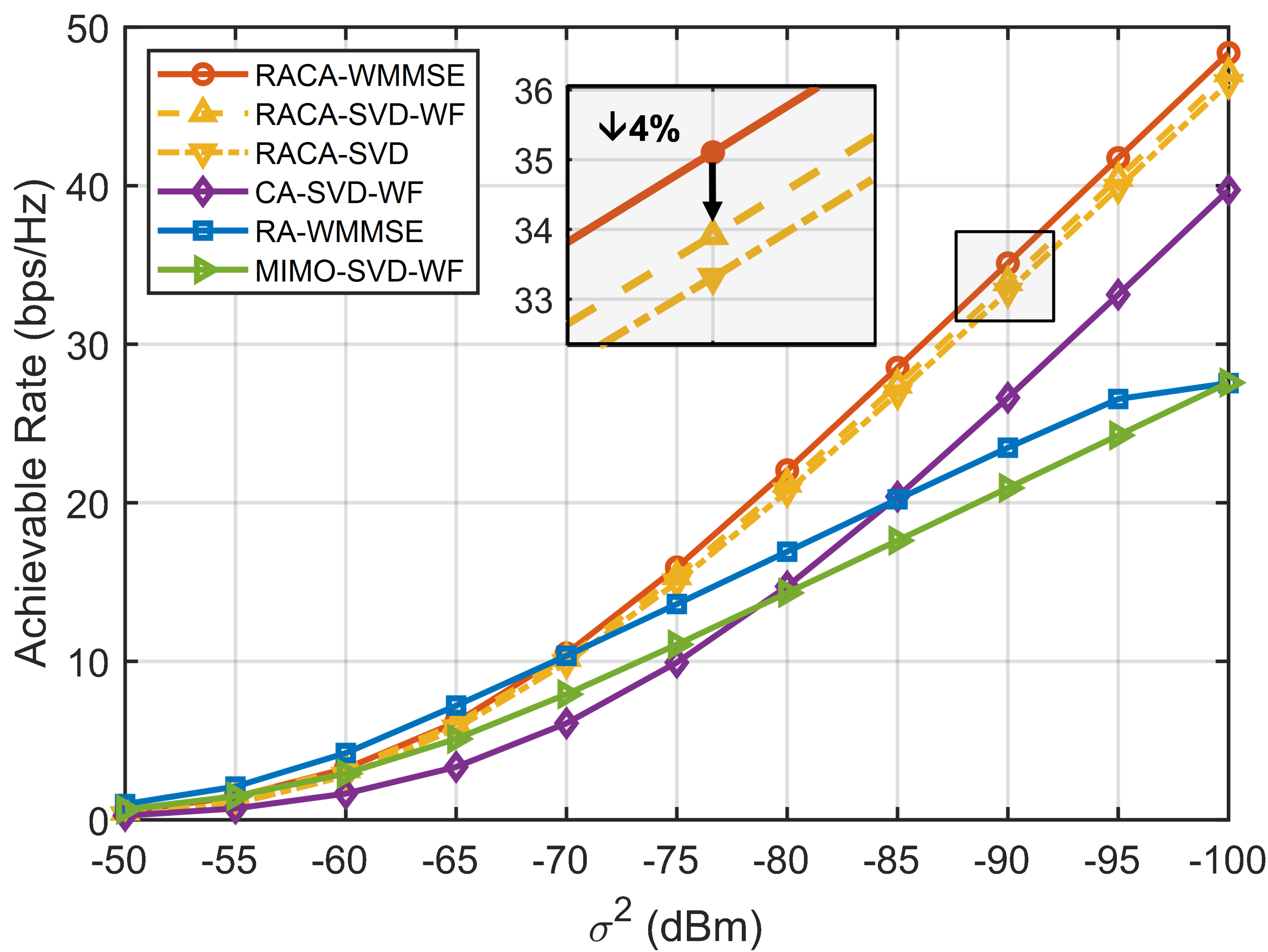}}
\caption{Rate of different MIMO schemes versus noise variance $\sigma^2$.}
\label{fig:rate}
\end{figure}

\begin{figure}[tbp]
\centerline{\includegraphics[width=0.85\linewidth]{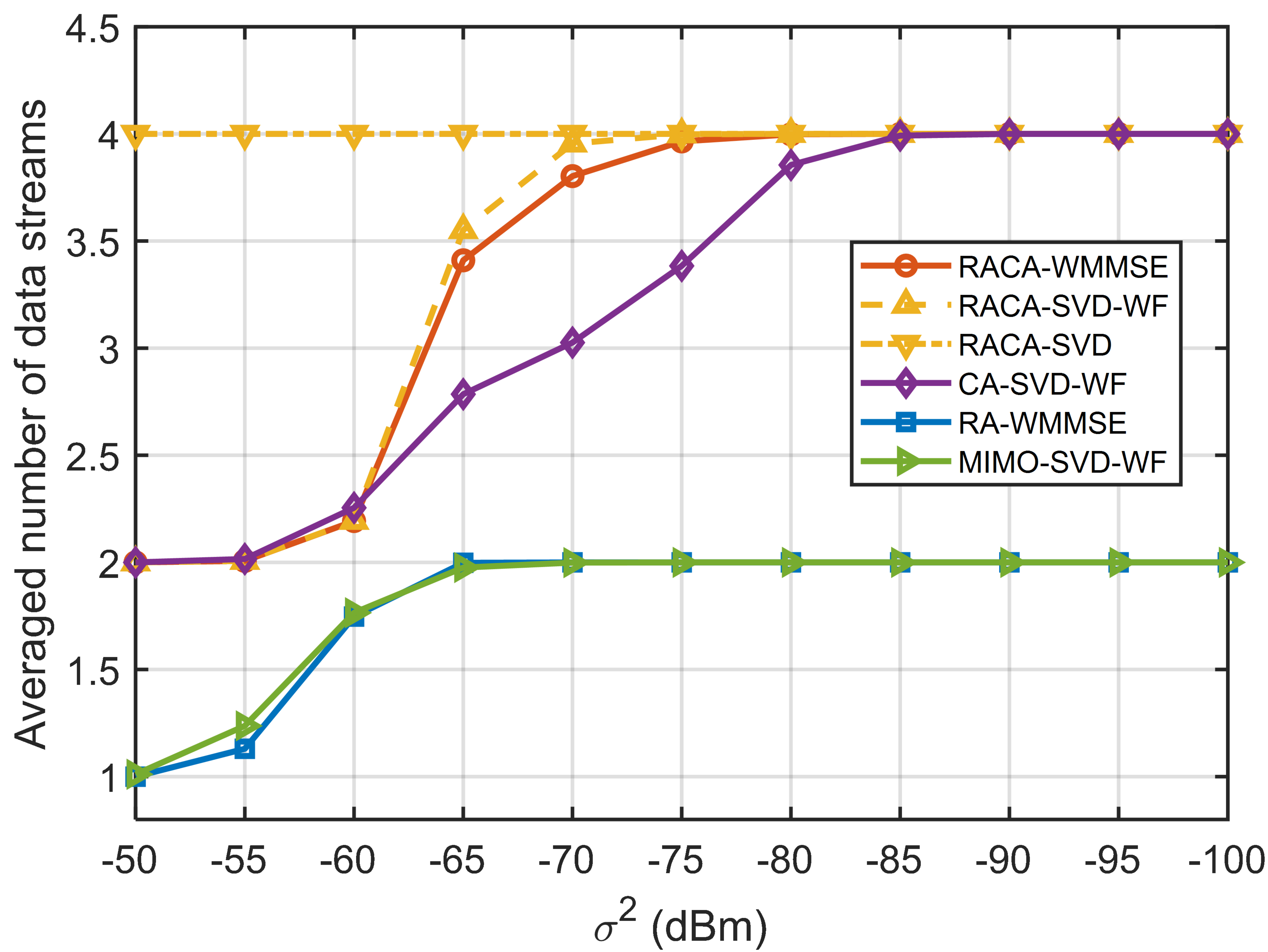}}
\caption{Number of transmitted data versus noise variance $\sigma^2$.}
\label{fig:stream}
\end{figure}

\subsection{Performance Comparison of Different MIMO Systems}
\label{ssec:convergence_rate}
Fig. \ref{fig:rate} and Fig. \ref{fig:stream} show the achievable rate and their corresponding number of transmitted data streams, respectively. Both are plotted against the noise variance, which effectively represents the SNR. 

In Fig. \ref{fig:rate}, it can be observed that our proposed \AlgoName{RACA-WMMSE/SVD(-WF)} and \AlgoName{CA-SVD-WF} have sharper slopes than \AlgoName{RA-WMMSE} and \AlgoName{MIMO-SVD-WF} due to the doubled data streams from $N_s$ to $2N_s$. Hence, the performance improvement benefitting from the MIMO scaling gain is significant with the increased SNR. As illustrated in Fig. \ref{fig:stream}, in the high SNR region, the number of transmitted data streams in the RACA and CA systems is twice as the other systems. Furthermore, the proposed \AlgoName{RACA-WMMSE/SVD(-WF)} are always superior to \AlgoName{CA-SVD-WF} owing to the lower signal attenuation. Specifically, its signal $\ve{s}^{f_H}$ only travels through a shorter $f_H$ path, then is translated and forwarded to a longer $f_L$ path by the relay in proximity. By contrast, \AlgoName{CA-SVD-WF} directly transmits $\ve{s}^{f_H}$ via a long $f_H$ path and suffers from a severe path loss. In summary, when $\sigma^2=-90$ dBm, the rate of \AlgoName{RACA-WMMSE} is improved by $32\%$ compared to that of \AlgoName{CA-SVD-WF}. Moreover, when $\sigma^2=-90$ dBm, the rate of \AlgoName{RACA-SVD-WF} achieves only $96\%$ of the rate of \AlgoName{RACA-WMMSE} because it does not jointly consider the antenna interference between the two links. Despite its slightly lower rate, the asynchronous channel updates in the distributed system with fewer signal exchanges make \AlgoName{RACA-SVD-WF} efficient and well-suited for fast-fading channels.


\begin{figure}[tbp]
\centerline{\includegraphics[width=0.85\linewidth]{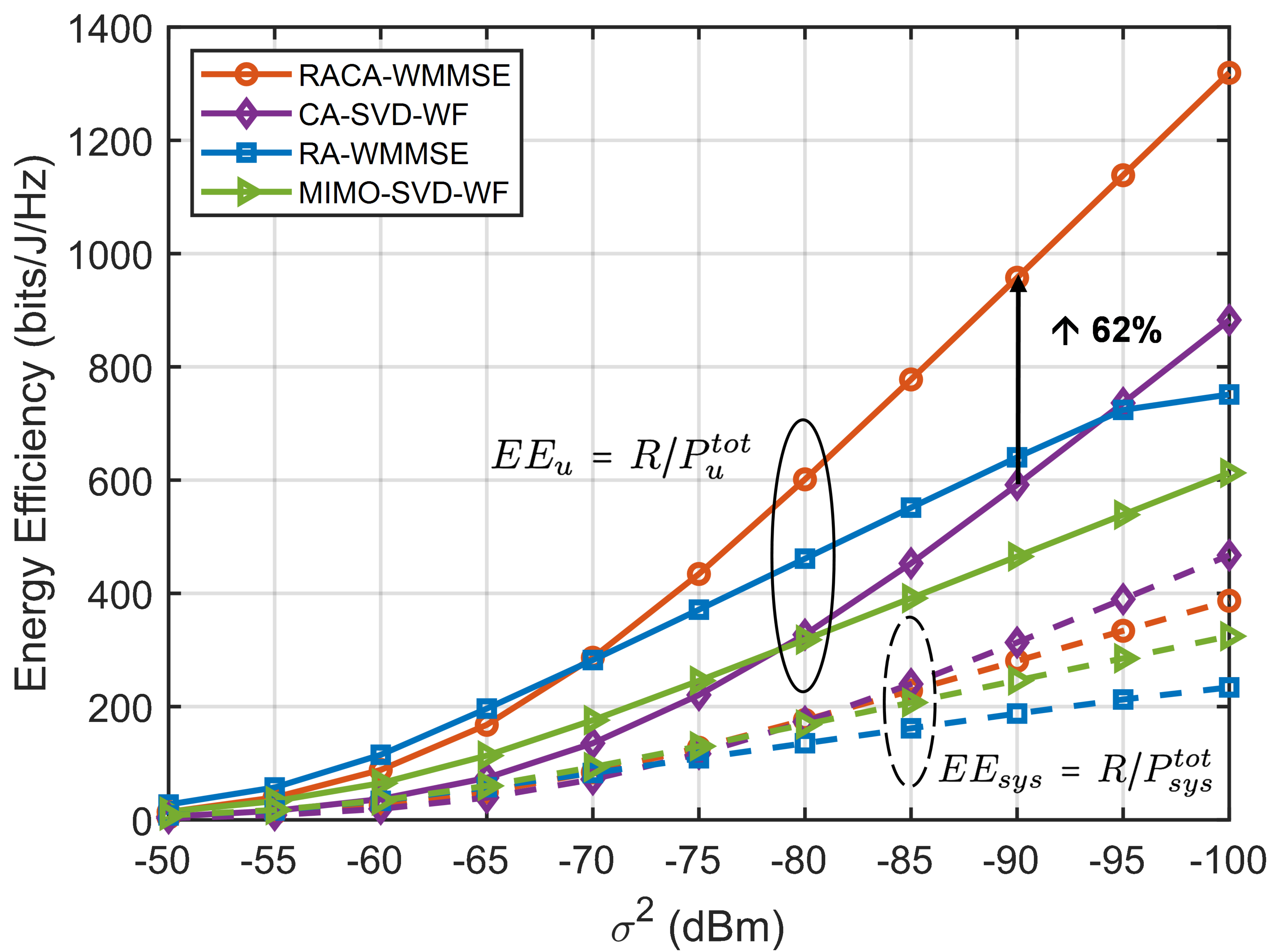}}
\caption{Energy efficiency of different MIMO schemes versus noise variance.}
\label{fig:EE}
\end{figure}

\begin{table}[tbp]
    \renewcommand\arraystretch{2.0}
    \caption{Parameters of Power Consumption Model}
    \begin{tabular}{|p{0.8cm}|p{4.2cm}|p{2.7cm}|}
        \hline
        \hspace{-0.3em}\textbf{System}   & \textbf{System's total power} $P_{sys}^{tot}$ & \textbf{UE's total power} $P_{u}^{tot}$ \\
        \hline
        \hspace{-0.2em}{RACA} & \hspace{-0.2em}$\frac{1}{\eta}(\bar{P}_{ua}^{f_L}+\bar{P}_{ur}^{f_H}+\bar{P_r})+(P^{c}_u+P^{c}_r+P^{c}_a)$&\hspace{-0.5em} $\frac{1}{\eta}(\bar{P}_{ua}^{f_L}+\bar{P}_{ur}^{f_H})+(P^{c}_u)$\\
        \hline
        \hspace{-0.2em}{\ \ CA} & \hspace{-0.2em}$\frac{1}{\eta}(\bar{P}_{ua}^{f_L}+\bar{P}_{ua}^{f_H})+(P^{c}_u+P^{c}_a)$ &\hspace{-0.5em} $\frac{1}{\eta}(\bar{P}_{ua}^{f_L}+\bar{P}_{ua}^{f_H})+(P^{c}_u)$\\
        \hline
        \hspace{-0.2em}{\ \ RA} & \hspace{-0.2em}$\frac{1}{\eta}(\bar{P}_{u}^{f_L}+\bar{P}_{r})+(P^{c}_u+P^{c}_r+P^{c}_a)$ &\hspace{-0.5em} $\frac{1}{\eta}(\bar{P}_{u}^{f_L})+(P^{c}_u)$\\
        \hline
        \hspace{-0.2em}{MIMO} & \hspace{-0.2em}$\frac{1}{\eta}(\bar{P}_{ua})+(P^{c}_u+P^{c}_a)$ &\hspace{-0.5em} $\frac{1}{\eta}(\bar{P}_{ua})+(P^{c}_u)$\\
        \hline
    \end{tabular}
    \label{table_EE}
    \footnotesize \\
    \footnotesize {$^1$ $\bar{P}$ is the actual transmitted power after optimization.\\} 
    \footnotesize {$^2$ $\eta=1.2$ is the power amplifier efficiency\cite{ref:power_model}. \\}
    \footnotesize {$^3$ $P^{c}_u=13 \ \text{dBm}, P^{c}_r=P^{c}_a=16 \ \text{dBm}$ are the circuit dissipated power at the UE, relay, and AP, respectively \cite{ref:power_model}.}
\end{table}

\subsection{Energy efficiency}
\label{ssec:energy_efficiency}
To ensure an immersive XR experience for users, lightweight XR devices must operate under stringent energy constraints due to their compact size, which limits the capacity for larger batteries. Consequently, energy efficiency (EE) emerges as another crucial metric in XR applications. We define two EE metrics to evaluate performances from both the system and the UE perspectives. $EE_{sys}=R/P_{sys}^{tot}$ captures the total power consumed across the entire system, including the static power overhead of relay circuits. On the other hand, $EE_{u}=R/P_{u}^{tot}$ specifically measures the power consumption of the UE, highlighting its impact on the battery life to maintain satisfactory performance. According to \cite{ref:power_model}, the detailed power consumption models for different systems are presented in Table \ref{table_EE}.

Fig. \ref{fig:EE} illustrates that $EE_{sys}$ of the RACA system is slightly lower than that of CA because the power overheads associated with the relay exceed the benefits of rate improvements facilitated by the relay assistance. Conversely, $EE_{u}$ of the RACA system shows a $62\%$ improvement over CA, benefiting from a combination of higher rates and reduced power consumption. By offloading power to the relay, energy efficiency is significantly enhanced, thus extending the battery life of XR devices and enhancing user experiences.

\subsection{Combinations of Frequency Bands}
\label{ssec:freq_band}
\begin{figure}[tbp]
\centerline{\includegraphics[width=0.85\linewidth]{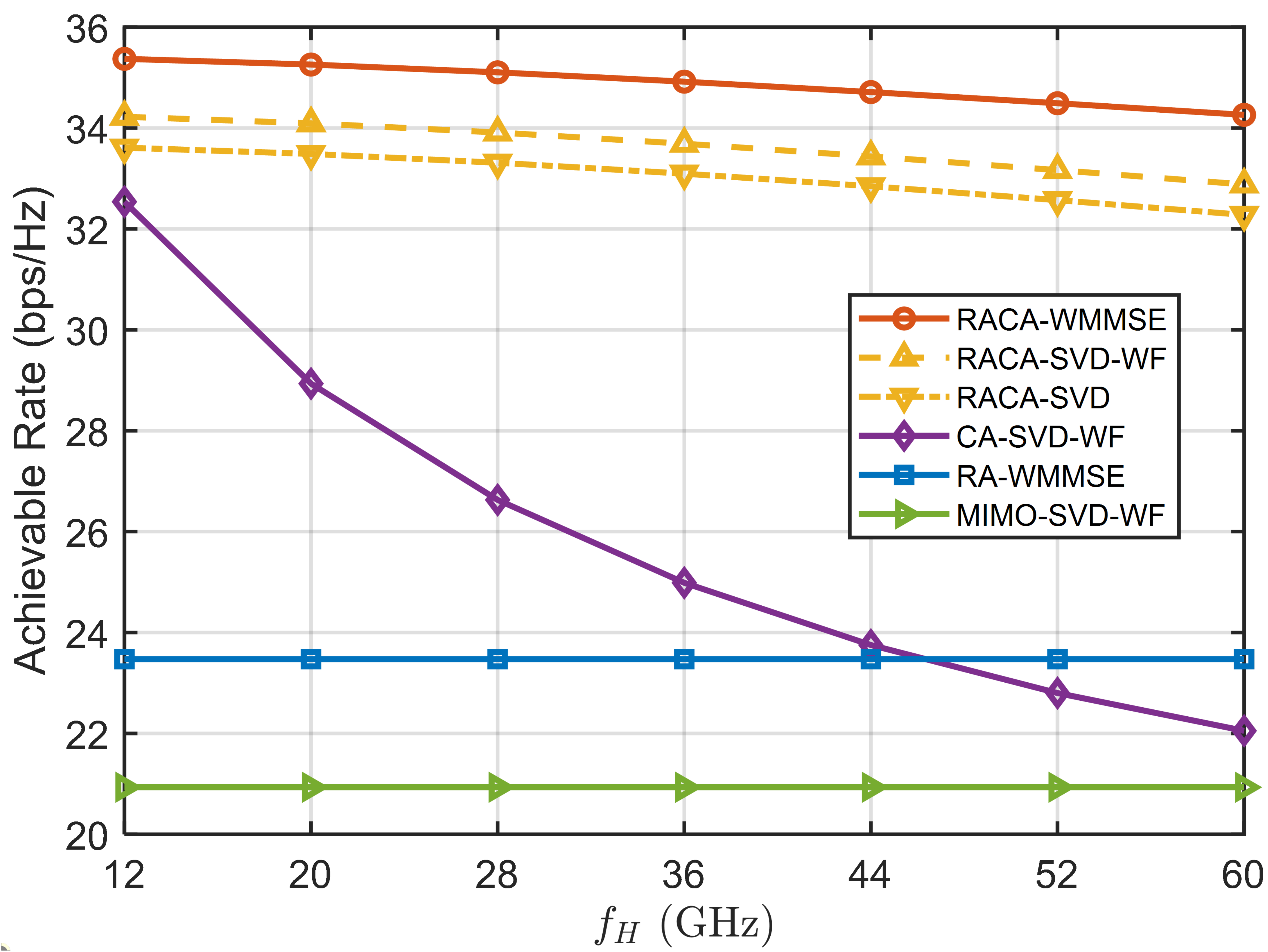}}
\caption{Rate versus different $f_H$ under $f_L=6$.}
\label{fig:freq_band}
\end{figure}

Given $f_L=6$ GHz, the results of different $f_H$ utilized in the systems are shown in Fig. \ref{fig:freq_band}. The rates of all frequency-division multiplexing systems decrease with the increased $f_H$ owing to the more severe attenuation. With lower $f_H$, \AlgoName{CA-SVD-WF} is superior to \AlgoName{RA-WMMSE} because of the multiplexing gain and shorter direct path. With higher $f_H$, despite limited rank, \AlgoName{RA-WMMSE} becomes better than \AlgoName{CA-SVD-WF} thanks to the boosted SNR in $f_L$ band. Furthermore, our proposed RACA combines both the advantages of CA and RA systems to achieve rank augmentation with better SNR. Therefore, the RACA system shows the efficiency in most frequency combinations to provide flexible spectrum usage.

\subsection{Power Allocation}
\label{ssec:power_allocation}
\begin{figure}[tbp]
\centerline{\includegraphics[width=0.85\linewidth]{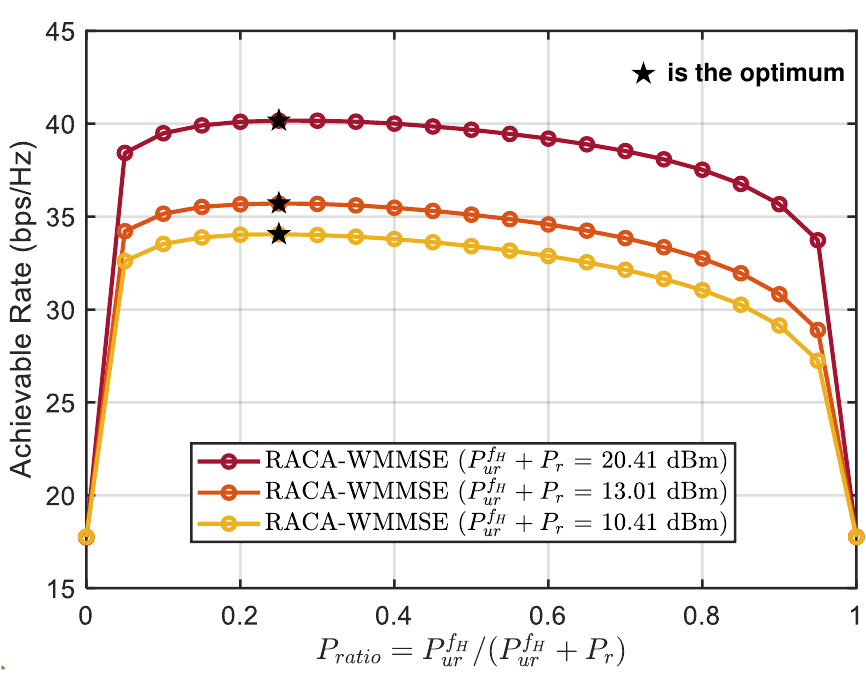}}
\caption{Rate of different power allocations between $\Pur$ and $P_r$.}
\label{fig:PA_TX_relay}
\end{figure}

In this subsection, the performance of various power allocations between $\Pur$ and $P_r$, as well as between $\Pua$ and $\Pur$ are illustrated. In Fig. \ref{fig:PA_TX_relay}, we first analyze the allocations between the UE and the relay with the same total power $\Pur+P_r$. We introduce $P_{ratio}^{UE-relay}=\Pur/(\Pur+P_r)\in [0,1]$ to indicate the proportion of power allocated at the UE. With lower $P_{ratio}^{UE-relay}$, the signal received at the relay is weaker, and the amplification gain of the relay amplifying matrix is higher, which results in a low received SNR at the relay and a high received SNR at the AP, and vice versa. It is observed that the system tends to balance the SNRs between the UE-relay and relay-AP links to avoid a transmission bottleneck in the two-hop links. Also, the optimal $P_{ratio}^{UE-relay}$ is insensitive to the total power $\Pur+P_r$, which indicates that the optimal SNRs of both the UE-relay and relay-AP links scale concurrently with the total power.

\begin{figure}[tbp]
\centerline{\includegraphics[width=0.85\linewidth]{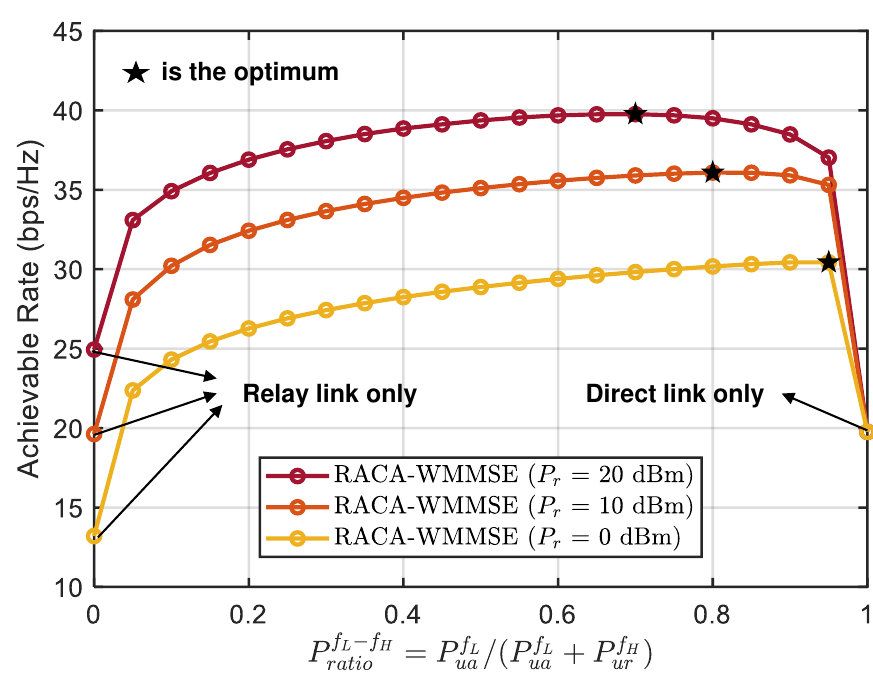}}
\caption{Rate of different power allocations between $\Pua$ and $\Pur$.}
\label{fig:PA_Fl_Fh}
\end{figure}

On the other hand, we define $P_{ratio}^{f_L-f_H}=\Pua/(\Pua+\Pur)\in [0,1]$ to indicate the proportion of power allocated at $f_L$ band, aiming to analyze the power allocations between two links at the UE. As shown in Fig. \ref{fig:PA_Fl_Fh}, when $P_{ratio}^{f_L-f_H} = 0$ or $1$, the system degrades as a relay link only or direct link only system, respectively. These two extreme cases cannot achieve rank augmentation. In addition, the optimal $P_{ratio}^{f_L-f_H}$ is variant to $P_r$. Because higher $P_r$ improves the SNR of the relay-AP link, the UE will tend to allocate more power at the $f_H$ band to enhance the SNR of the UE-relay link. In contrast, the UE tends to allocate more power to the shorter direct link rather than the weak relay link bottlenecked by the lower $P_r$.

\section{Conclusion}
\label{sec:conclusion}
In this paper, the RACA system is investigated to enhance the rate of a lightweight XR device with the nearby relay collaboration. The advantages of the frequency-division multiplexing in the CA system and the SNR enhancement in the RA system are simultaneously leveraged in the RACA system. For different scenarios, we propose two transmission protocols and their respective optimization algorithms to tackle the non-convex rate maximization problem. In a centralized system, the convergence-guaranteed WMMSE jointly mitigates the antenna interference. In a distributed system, the rate of the low-complexity SVD-WF achieves $96\%$ of the rate of WMMSE and supports asynchronous channel updates. Simulation results demonstrate that the rate of the RACA system achieves $32\%$ and $50\%$ improvements compared to the rate of the CA and RA systems, respectively, which is suitable for the high data rate-required XR applications on limited-resource XR devices.

\section*{Appendix A: Proof of Theorem 1} \label{sec::appendix_A}
\begin{proof}
Suppose that $\{\Wa^*, \Wua^*, \Wur^*, \ma{\Psi}^*, \ma{Z}^*\}$ is an optimal solution for problem \eqref{P_wmmse}, then its optimal MSE matrix can be derived as
\begin{align}
\ma{E}^* & \stackrel{(a)}{=} \ma{I}_{2N_s}-\Wa^*\ma{H}^*-\ma{H}^{*H}\Wa^{*H} \notag\\
& \quad \ + \Wa^*(\ma{H}^*\ma{H}^{*H}+\ma{J}^*)\Wa^{*H} \notag\\
& \stackrel{(b)}{=} \ma{I}_{2N_s}-\ma{H}^{*H}(\ma{H}^*\ma{H}^{*H}+\ma{J}^*)\ma{H}^* \notag\\
& \stackrel{(c)}{=} (\ma{I}_{2N_s}+\ma{H}^{*H}\ma{J}^{*-1}\ma{H}^*)^{-1},
\end{align}
where $\ma{H}^*$ and $\ma{J}^*$ are the optimal effective channel and noise covariance matrix with substitution of the optimal solution. $(a)$ and $(b)$ hold due to the expansion of \eqref{MSE_matrix} and the substitution of \eqref{Wa_opt}, respectively. According to the matrix inversion lemma \cite{ref:boyd2004convex}, $(c)$ holds. Therefore, $\ma{Z}^*$ can be explicitly given as
\begin{equation}
    \ma{Z}^*=(\ma{E}^*)^{-1}=\ma{I}_{2N_s}+\ma{H}^{*H}\ma{J}^{*-1}\ma{H}^*.
    \label{Z_global_opt}
\end{equation}
Then, with substitution of \eqref{Z_global_opt} into \eqref{P_wmmse_min}, the minimized \eqref{P_wmmse_min} is equivalent to 
\begin{equation}
    \hspace{-2em}\underset{\Wua^*, \Wur^*, \ma{\Psi}^*} {\text{max}}\text{log}|\ma{I}_{2N_s}+\ma{H}^{*H}\ma{J}^{*-1}\ma{H}^*|,
\end{equation}
which maximizes the rate expressed in \eqref{P_rate_max}. Hence, the global optimum for the WMMSE problem \eqref{P_wmmse} is identical to the global optimum for the rate maximization problem \eqref{P_rate}.
\end{proof}

\bibliographystyle{IEEEtran}
\bibliography{refs}

\begin{IEEEbiography}[{\includegraphics[width=1in,height=1.25in,clip,keepaspectratio]{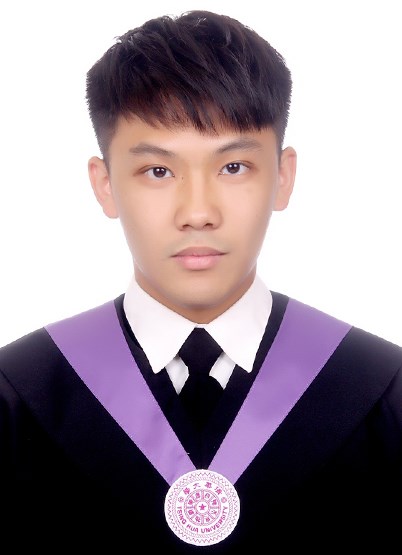}}]{Chi-Wei Chen}
(S’23) received his B.S. degree in electrical engineering from National Tsing Hua University, Hsinchu, Taiwan, in 2020. He is currently pursuing a Ph.D. degree in the Graduate Institute of Electronics Engineering, National Taiwan University. His research interests are in the areas of cooperative wireless communication, IRS-assisted wireless communication systems design, and VLSI architecture for DSP.
\end{IEEEbiography}
\begin{IEEEbiography}
[{\includegraphics[width=1in,height=1.25in,clip,keepaspectratio]{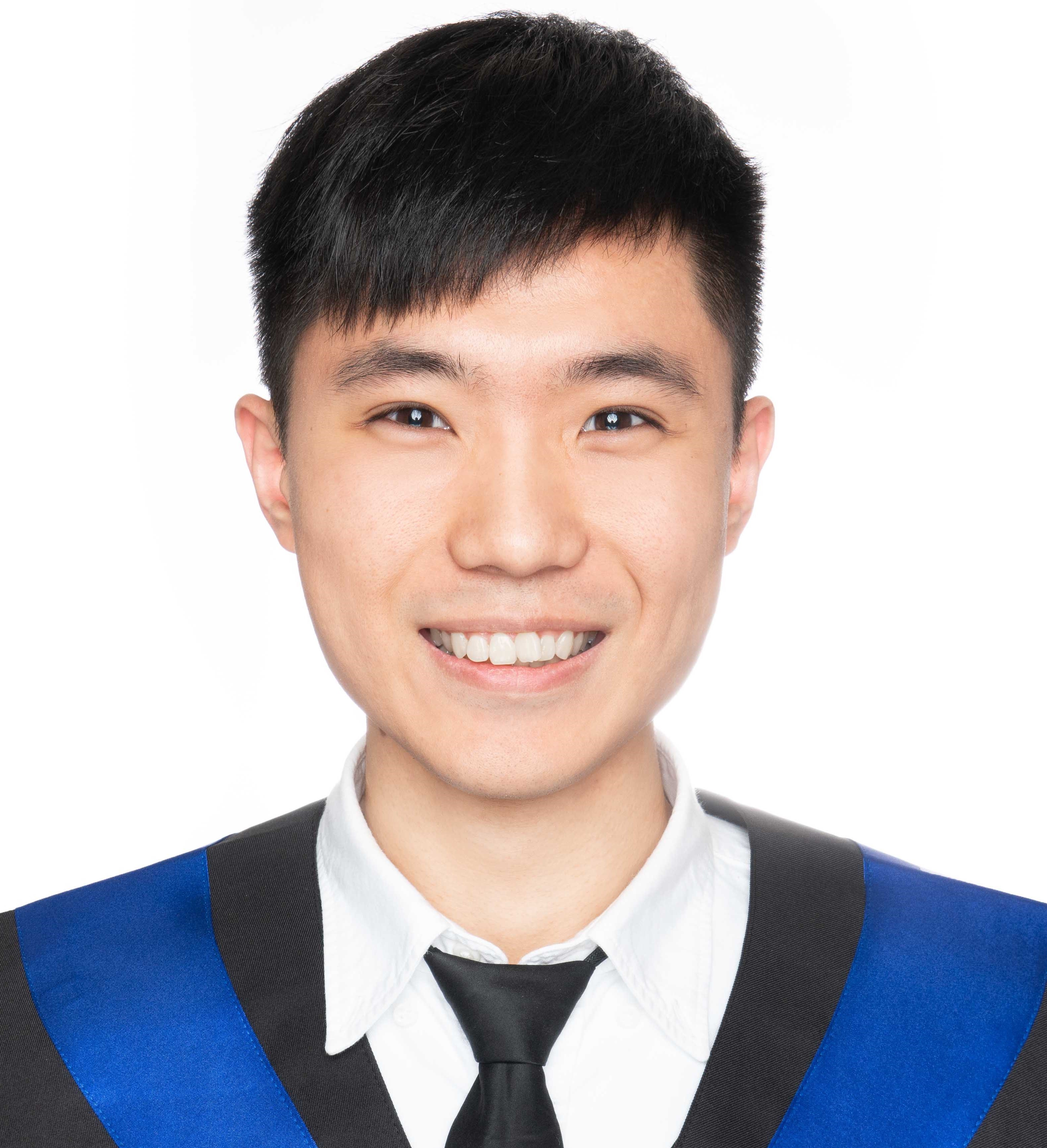}}]{Wen-Chiao Tsai}
(S’23) received the B.S. degree in electrical engineering from National Taiwan University, Taipei, Taiwan, in 2020, where he is currently pursuing the Ph.D. degree with the Graduate Institute of Electronics Engineering. His research interests are in the areas of machine learning-assisted wireless communication systems design, IRS-assisted wireless communication systems design, and VLSI architecture for DSP.
\end{IEEEbiography}

\begin{IEEEbiography}
[{\includegraphics[width=1in,height=1.25in,clip,keepaspectratio]{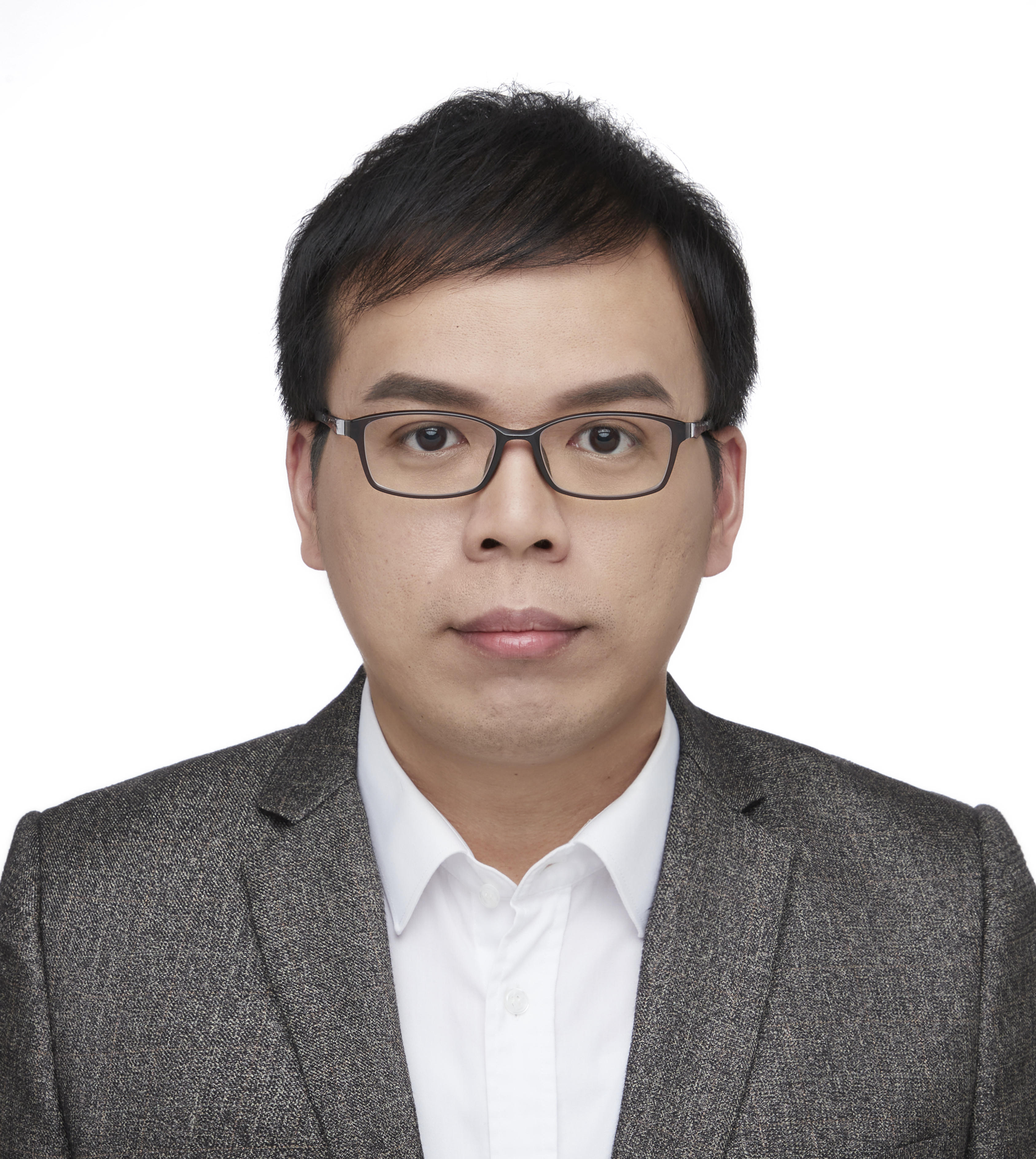}}]{Lung-Sheng Tsai}
Lung-Sheng Tsai received the B.S. degree in electrical engineering from the National Tsing Hua University, Taiwan, in 2002, the M.S. degree in communication engineering from National Chiao Tung University, Taiwan, in 2004, and the Ph.D. degree in Graduate Institute of Communication Engineering from National Taiwan University, in 2010. Currently, he is a Technical Manager at MediaTek Inc., focusing his research on MIMO systems and the development of next-generation wireless networks.
\end{IEEEbiography}

\begin{IEEEbiography}
[{\includegraphics[width=1in,height=1.25in,clip,keepaspectratio]{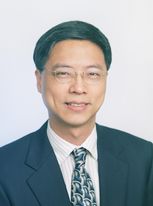}}]{An-Yeu (Andy) Wu}
(M’96-SM’12-F’15) received the B.S. degree from NTU in 1987 and the M.S. and Ph.D. degrees from the University of Maryland, College Park, in 1992 and 1995, respectively, all in electrical engineering. In August 2000, he joined as a Faculty Member of the Department of Electrical Engineering and the Graduate Institute of Electronics Engineering, NTU, where he is currently a Distinguished Professor. His research interests include low-power/high-performance VLSI architectures and IC designs for DSP/communication/AI applications and adaptive/bio-medical signal processing. From 2020 to 2021, he served as the Editor-in-Chief (EiC) of EEE Journal on Emerging and Selected Topics in Circuits and Systems (JETCAS).
\end{IEEEbiography}

\end{document}